\documentclass[a4,12pt, twocolum,  usenatbib]{mn2e}
\usepackage[dvipdf]{graphicx, psfrag}
\usepackage{times}
\usepackage{color}
\usepackage{amsmath, amssymb}
\usepackage{ulem}

\usepackage{natbib}
\usepackage{aas_macros}

 \title[A numerical method for baroclinic stars]
 {A versatile numerical method for obtaining structures of rapidly rotating baroclinic stars:
 self-consistent and systematic solutions with shellular-type rotation
}
\author[K. Fujisawa]
{Kotaro Fujisawa
\thanks{E-mail: fujisawa@heap.phys.waseda.ac.jp}
\\
Advanced Research Institute for Science and Engineering, 
Waseda University, 3-4-1 Okubo, Shinjuku-ku, Tokyo 169-8555, Japan \\
}
\date{Accepted 2015 September 17. Received 2015 Sep 17; in original form 2015 July 9}

\def\Vec#1{\mbox{\boldmath $#1$}}

\def\P#1#2{\dfrac{\partial #1}{\partial #2}}

\begin{document}

\maketitle

\begin{abstract}
This paper develops a novel numerical method for obtaining structures 
of rapidly rotating stars based on a self-consistent field scheme.
The solution is obtained iteratively.
Both rapidly rotating barotropic and baroclinic equilibrium 
states are calculated self-consistently using this method.
Two types of rotating baroclinic stars are investigated
by changing the isentropic surfaces inside the star.
Solution sequences of these are 
calculated systematically and critical rotation models
beyond which no rotating equilibrium state exists are also obtained.
All of these rotating baroclinic stars satisfy necessarily the Bjerknes--Rosseland rules.
Self-consistent solutions of baroclinic stars with shellular-type 
rotation are successfully obtained where the isentropic surfaces are oblate and 
the surface temperature is hotter at the poles than at the equator 
if it is assumed that the star is an ideal gas star.
These are the first self-consistent and systematic solutions of
rapidly rotating baroclinic stars with shellular-type rotations.
Since they satisfy the stability criterion due to their rapid rotation,
these rotating baroclinic stars would be dynamically stable. 
This novel numerical method and the solutions of the rapidly 
rotating baroclinic stars will be useful for investigating 
stellar evolution with rapid rotations. 
\end{abstract}

\section{Introduction}

One of the most fascinating challenge in the stellar 
astrophysics is understanding the 
structures of realistic rotating stars and their  
evolution (for example, see \citealt{Maeder_Meynet_2000}; \citealt{Langer_2012}). 
According to recent observations, many stars display detectable rotations.
In fact, massive  stars in particular, such as B and O stars, 
tend to have clearly rapid rotation 
(e.g. \citealt{Hunter_et_al_2008}).
Approximately 10 per cent of them have the rotational velocities which are
larger than 300 $\mathrm{km\,s^{-1}}$ (\citealt{Ramirez-Agudelo_et_al_2013}; \citealt{Dufton_et_al_2013}).
Such rapid rotational velocities can reach
the critical velocity beyond which mass shedding due to rotation
occurs and no equilibrium state exists (\citealt{Maeder_Book}),
because the typical values of the critical velocities 
for 20 $M_\odot$ star with solar metallicity
are $\sim$ 300 $\mathrm{km\,s^{-1}}$ - 600 $\mathrm{km\,s^{-1}}$ during the main sequence phase
 (\citealt{Ekstrom_et_al_2008}).
The shape of these rapidly rotating stars is oblate due to
the centrifugal force. Therefore, 
their outer layers cannot be well described by spherical models 
due to their fast rotation and so non-spherical models are required 
to describe them. The structures of  realistic rotating stars are described
by using many realistic physical properties and processes such as viscosity,
inhomogeneous chemical composition and meridional flows (not considered in this paper).
In general, the surfaces of equal density (isopycnic surfaces) and equal pressure (isobaric surfaces) 
do not coincide due to these physical properties and processes,
i.e. the stars  becomes baroclinic.

\begin{figure*}
\includegraphics[width=8.2cm]{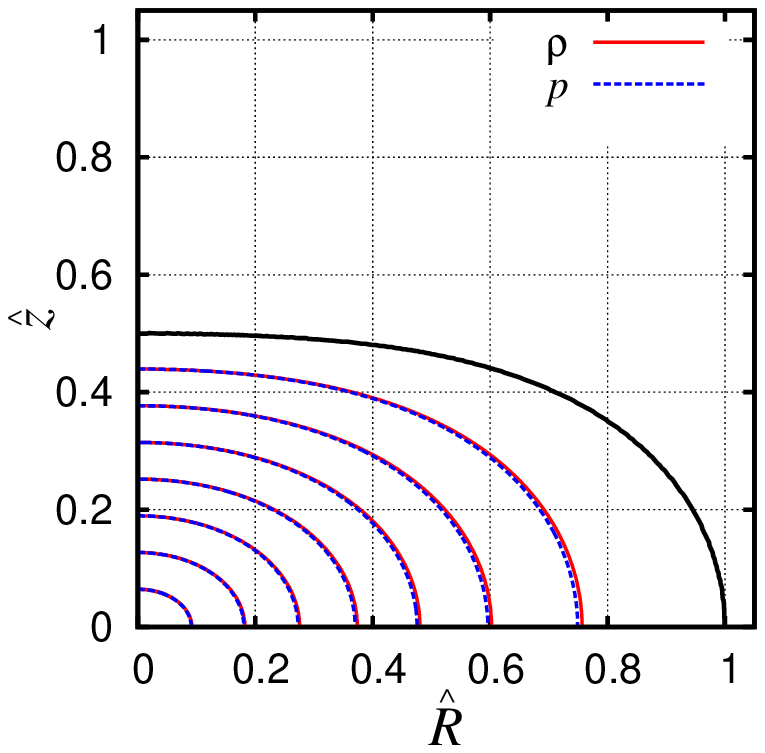}
 \includegraphics[width=8.2cm]{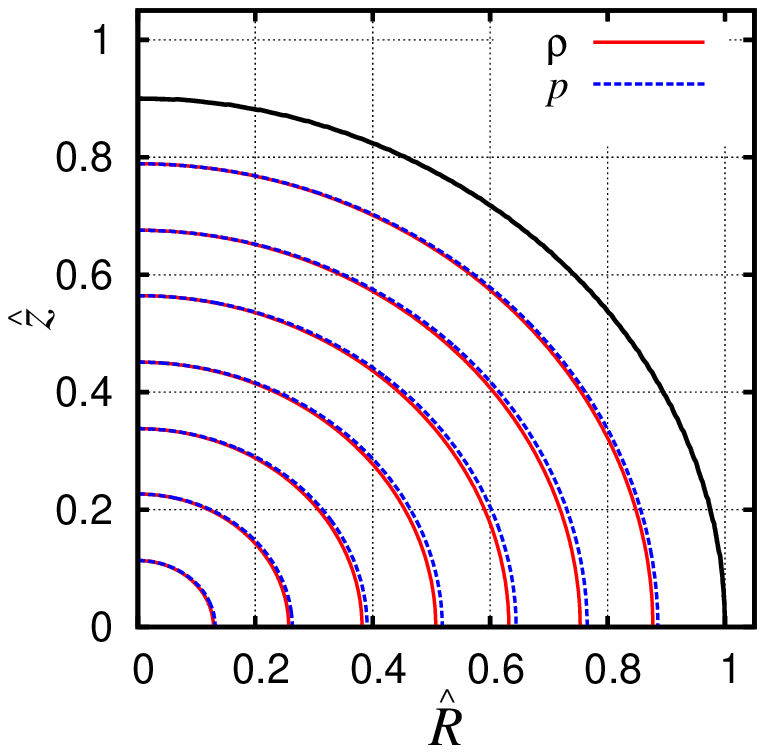}
\caption{Seven arbitrary isopycnic surfaces (solid curves) and isobaric surfaces (dashed curves) 
of baroclinic rotating stars. The outermost curves denote the stellar surface. The isopycnic surfaces are more oblate than
the isobaric surfaces in the left-hand  panel, while the isobaric surfaces are more oblate than the isopycnic surfaces
in the right-hand panel. These two models are different classes of baroclinic stars.}
\label{Fig:iso_surfaces}
\end{figure*}

Fig. \ref{Fig:iso_surfaces} displays two examples of rapidly rotating baroclinic stars. 
The isopycnic surfaces (solid curves) are inclined to the isobaric surfaces (dashed curves). 
The isopycnic surfaces are more oblate than the isobaric 
surfaces in the left model, while the isobaric surfaces are more
oblate than the isopycnic surfaces in the right model.
Therefore, the value of the density on the isobaric surface varies in the baroclinic star.
On each isopycnic surface, the density in the left model increases from the pole to the equator,
while the density in the right model decreases from the pole to the equator.
These two models are different classes of baroclinic stars 
explained in detail in Sec. 2.1.

The baroclinicity is characterized 
by the angle between isopycnic surfaces and isobaric surfaces. 
It is assumed that the system is  stationary and axisymmetric and 
the star does not have magnetic field and meridional flow, 
the Euler equation of the rotating star is 
\begin{eqnarray}
 \frac{1}{\rho}\nabla p = - \nabla \phi + R \Omega^2 \Vec{e}_R,
\end{eqnarray}
where $\rho$, $p$, $\phi$ and $\Omega$ are, respectively, the density, the pressure,
the gravitational potential and the angular velocity of the star.  
Cylindrical coordinates ($R$, $\varphi$, $z$) are utilized in this equation. 
The $\varphi$ component of the curl of this  equation is 
 \begin{eqnarray}
 \frac{1}{\rho^2} \left(\nabla p \times \nabla \rho \right)_\varphi
 = \P{R \Omega^2}{z}.
\label{Eq:rot_Euler}
 \end{eqnarray}
If the isopycnic surfaces and the isobaric surfaces coincide, i.e.
the pressure depends on the density only, the star is barotropic ($p = p(\rho)$)
and the left-hand side of the equation vanishes.
The rotation of the barotropic star is rigid ($\Omega = \Omega_0$, constant) or cylindrical
($\Omega = \Omega(R)$) because the centrifugal force can be 
derived from a rotational potential (e.g. \citealt{Eriguchi_Muller_1985}; \citealt{Hachisu_1986a,Hachisu_1986b};
\citealt{Maeder_Book}; \citealt{Fujisawa_2015a}) as 
 \begin{eqnarray}
0 = \P{R \Omega^2}{z} \Rightarrow \nabla \phi_{\mathrm rot.} = R \Omega^2(R) \Vec{e}_R,
\label{Eq:rot_pot}
 \end{eqnarray}
where $\phi_{\mathrm{rot.}}$ is the rotational potential that is an arbitrary function of $R$.
If the isopycnic surfaces are inclined to the isobaric surfaces, 
the left-hand side of equation (\ref{Eq:rot_Euler}) does not vanish.
 The angular velocity distributions 
in the baroclinic stars are no longer constant and cylindrical due to the differences between 
isopycnic surfaces and isobaric surfaces.
The situation is completely different from baroclinic star.
Therefore, baroclinicity is an important component to consider in order to 
investigate such realistic rotating stars and their evolution.

Nevertheless, only a few works have succeeded in obtaining baroclinic 
equilibrium states with rapid rotation self-consistently
because it is difficult to calculate and analyse baroclinic stars.
If it is assumed that the star is barotropic, 
the rotation becomes cylindrical and the centrifugal force can be derived from the 
potential in equation (\ref{Eq:rot_pot}). Therefore, the first integral of the Euler equation
can be obtained analytically as
\begin{eqnarray}
 \int \frac{dp}{\rho} = - \phi + \phi_{\mathrm {rot.}} + C,
\label{Eq:rot_phi}
\end{eqnarray}
where $C$ is the integral constant.
Since the system is barotropic ($p = p(\rho)$), 
the integral of the left-hand side of this equation can be 
also calculated. This first integral of the Euler equation is used by many 
previous works for obtaining rapidly rotating barotropic stars  
(e.g. \citealt{Eriguchi_Muller_1985}; \citealt{Hachisu_1986a, Hachisu_1986b}).
In contrast, the situation of the baroclinic star is more complicated. 
If the star is baroclinic, the first integral of the 
Euler equation cannot be obtained analytically 
because the centrifugal force cannot be derived from a rotational potential
and the distributions of the angular velocity must be calculated self-consistently 
by solving equation (\ref{Eq:rot_Euler}). 

In order to avoid this difficulty,
many previous works treated rotating baroclinic stars as slowly rotating stars by using 
perturbative methods (\citealt{Roxburgh_Griffith_Sweet_1965}; \citealt{Roxburgh_Strittmatter_1966};
\citealt{Faulkner_Roxburgh_Strittmatter_1968}; \citealt{Clement_1969}; \citealt{Sackmann_Anand_1970}; 
\citealt{Kippenhahn_Thomas_1970}; \citealt{Monaghan_1971}; \citealt{Sharp_Smith_Moss_1977}). 
Others calculated self-consistent rotating equilibrium states, 
but treated rotating stars as pseudo-barotropic, i.e. 
isobaric surfaces and isopycnic surfaces are not inclined
(\citealt{Jackson_1970}; \citealt{Papaloizou_Whelan_1973};
\citealt{Eriguchi_Mueller_1991}; \citealt{Jackson_et_al_2005}).
Although these studies considered the thermal balance of a 
rotating star in these works, the isothermal surfaces 
are also isobaric and so the system is essentially barotropic.
Only a few works succeeded in obtaining rapidly rotating 
baroclinic stars self-consistently.

\cite{Uryu_Eriguchi_1994} investigated 
self-consistent rotating baroclinic  equilibrium states.
They discretized all basic equations of fully radiative stars and 
calculated them simultaneously by using Straight-Forward 
Newton-Raphson scheme (\citealt{Eriguchi_Muller_1985}).
The numerical code was extended by \cite{Uryu_Eriguchi_1995}
and  more realistic stars with a convective core and radiative envelope were calculated.
\cite{Roxburgh_2006} obtained self-consistent rotating baroclinic stars
with shellular rotation using an iterative method (\citealt{Roxburgh_2004}).
\cite{Roxburgh_2006} did not adopt a particular equation of state and 
did not solve a thermal balance equation.
Instead, both density and pressure distributions were obtained independently and 
calculated iteratively by fitting the radial profile of the
angular velocity. \cite{Espinosa_Rieutord_2007} succeeded in obtaining self-consistent
rotating baroclinic stars with viscosity and meridional flows  
by using a pseudo-spectral method (cf. \citealt{Rieutord_1987}). 
In their numerical code, the physical variables 
are projected on to the spherical harmonics and solved 
iteratively via a  generalized algorithm for polytropic stellar models (\citealt{Bonazzola_et_al_1998}).
More recently, the  numerical code was improved 
and many realistic and complex solutions were obtained by \cite{Espinosa_Rieutord_2013}.
Very recently, \cite{Yasutake_Fujisawa_Yamada_2015} developed 
a new formulation and numerical method based on the Lagrangian variational principle
to obtain rotating self-gravitating configurations.
They were able to obtain both rotating barotropic 
and baroclinic equilibrium sates consistently  by using their new method. 

These studies have improved self-consistent rotating baroclinic models, 
but the basic understanding of rotating baroclinic  stars is still unclear
because there are only the few solutions of such stars that were developed in
the aforementioned works. 
Many solution sequences and systematic studies are required to 
improve  the understanding of rotating baroclinic stars and their evolution. 
Simple and powerful numerical methods are needed to systematically obtain 
solution sequences. As a first step towards 
realistic rotating  stars and their evolution, 
this paper investigates a versatile  numerical method for 
systematically obtaining structures of rapidly rotating baroclinic stars
as well as solutions for baroclinic rotating stars.
A simple equation of state is adopted in this paper
in order to obtain rotating baroclinic stars easily and systematically.
Some previous self-consistent works adopted 
complex thermal balance equations, but the rotating baroclinic 
equilibrium state itself is determined without any 
explicit thermal balance equation
(\citealt{Roxburgh_2006}; \citealt{Yasutake_Fujisawa_Yamada_2015}).
The mechanical structures themselves are determined independently
from assumed or additional thermal structures 
which are totally independently and freely introduced to 
mechanical structures in Newtonian gravity.
It should be noted that the mechanical part and the thermal part
cannot be separated in general relativistic situations 
because the total energy density 
is one of the source of the gravity in the Einstein equations 
(\citealt{Yoshida_Kiuchi_Shibata_2012}).
Therefore, in the Newtonian gravity framework, 
we can obtain rotating baroclinic stars systematically 
even when we ignore the thermal balance equation.
This paper adopts a simple equation of state 
as a baroclinic equation of state,
but a realistic equation of state could also be easily integrated into the method. 
This new method can calculate rapidly rotating baroclinic stars easily.  

The remainder of this paper is organized as follows. 
Section 2 describes the formulation 
of the rotating baroclinic star 
and numerical scheme of the new method.
Section 3.1 verifies the accuracy of the numerical results and
compares them  with results obtained via another numerical scheme.
The numerical results and solution sequences are displayed 
in Section 3.2. Discussion is given in Sections 4.1 and 4.2. 
The paper is summarized in Section 4.3.

\section{Formulation and  numerical method}

The formulation and details of the numerical method are 
described in this section.
The formulation of rotating baroclinic 
stars is summarized in Section 2.1. 
The remainder of this section focuses on
explaining the new numerical scheme briefly.
Both spherical polar coordinates $(r, \theta, \varphi)$ and 
cylindrical coordinates $(R, \varphi, z)$ are utilized 
in the formulation, but spherical polar 
coordinates are employed in the actual numerical code. 

\subsection{A formulation of rotating baroclinic stars}

It is assumed that the system is in a stationary state and the configuration 
is axisymmetric about the rotational axis. 
It is also assumed that the configuration is equatorially symmetric. 
Both meridional flow and magnetic field are neglected in this paper.
Although the meridional flow plays important role 
for secular transport processes in stellar interiors (e.g. \citealt{Mathis_LNP_2013}), 
the typical time-scale of the meridional flow is much longer than the 
dynamical time-scale of the star. It is almost negligible 
to consider the mechanical structures of the rapidly rotating star in equilibrium.
Under these assumptions, the basic equations that describe the 
rotating baroclinic star are the Euler equations  
\begin{eqnarray}
 \frac{1}{\rho} \P{p}{r} = - \P{\phi}{r} + r \sin^2 \theta \Omega^2,
\label{Eq:eular_r}
\end{eqnarray}
\begin{eqnarray}
 \frac{1}{\rho}\P{p}{\theta} = - \P{\phi}{\theta} + r^2 \sin \theta \cos \theta \Omega^2,
\label{Eq:eular_th}
\end{eqnarray}
the integral form of the Poisson equation 
\begin{eqnarray}
  \phi = - G \int \frac{\rho (\Vec{r}')}{|\Vec{r} - \Vec{r}'|} d^3 \Vec{r}',
   \label{Eq:Poisson_int}
\end{eqnarray}
and a baroclinic equation of state
\begin{eqnarray}
 p = p(\rho, X),
\end{eqnarray}
where $\rho$, $p$, $\phi$, $\Omega$, $G$, and $X$ are, respectively, 
the density, the pressure, the gravitational potential, the angular velocity,
the gravitational constant and  a scalar function that characterizes 
a baroclinicity such as temperature $T$ or specific entropy $s$.
In this paper, the following analytic form of a simple equation of state is adopted:
\begin{eqnarray}
&& p(\rho, K) = K(r, \theta) \rho^{1 + 1/N}, \nonumber \\
&& K(r,\theta) = K_0 
\left\{ 1 + \epsilon \left( \frac{\sin^2 \theta }{a_0^2} + \frac{\cos^2 \theta}{b_0^2} \right) r^m \right\},
\label{Eq:p_k_rho}
\end{eqnarray}
where $\epsilon$, $a_0$, $b_0$, $K_0$, $m$ are parameters.
$K$ is related to the specific entropy of the star.  
If we assume the ideal gas star ($p \sim T \rho$), then $K$ is proportional 
to the specific entropy ($K \propto \exp (s) $) and
the distribution of $K$ represents the distribution of the specific entropy.
If the $K$ at the equator is lager than at the pole,
the surface temperature is hotter at the equator than at the pole.
If the $K$ at the equator is smaller than that at the pole
the surface temperature is cooler at the equator than at the pole.
Although this equation of state is simple,
two types of rotating baroclinic stars in Fig. \ref{Fig:iso_surfaces} are obtained
by this equation of state.
Spherical isentropic (constant entropy) and oblate isentropic 
models were investigated by fixing the
values of $a_0$ and $b_0$ of this analytic form of $K$, noting that 
the isentropic surfaces becomes spherical when $a_0 = b_0$ and oblate when $a_0 > b_0$ respectively.
The spherical isentropic model has the spherical isentropic surfaces and 
the oblate isentropic models has the oblate isentropic surfaces. These two types of 
isentropic models become two types of rotating baroclinic models in Fig. \ref{Fig:iso_surfaces}.
The value of $\epsilon$ denotes the degree of baroclinicity. If we fix $\epsilon = 0$, 
the system becomes barotropic since $K$ becomes 
constant and $p$ is a function of $\rho$ $(p = p(\rho))$. 
However, if we choose $\epsilon \neq 0$, then 
$K$ is no longer constant and the system becomes baroclinic. 
As the absolute value of $\epsilon$ increases, 
the baroclinicity of the star also increases. 
As we will see in Section 3, this simple equation of state systematically
 provides  various rotating baroclinic stars 
by fixing the parameters in this ways.
Since the isentropic surfaces are oblate in the oblate isentropic model,
the $K$ at the equator is smaller than at the pole.
Therefore, the surface temperature of the oblate isentropic model
is hotter at the pole than at the equator. 
From a observational point of view, the surface temperatures of the 
rapidly rotating early type stars are hotter at the poles than that at the equator 
(e.g. $\alpha$ Leonis observed by \citealt{Che_et_al_2011}) by the gravitational
darkening (\citealt{von_Zeipel_1924}). Therefore,
the oblate isentropic model might be favoured in massive stars,
but both spherical and oblate isentropic models were calculated 
to obtain various rotating baroclinic stars systematically in this paper.

The $\varphi$ component of the curl of the Euler equation is used
in this numerical method instead of the $\theta$ component of the Euler equation 
(equation \ref{Eq:eular_th}), i.e.,
\begin{eqnarray}
 r^2 \sin \theta \cos \theta \P{\Omega^2}{r} 
- r\sin^2 \theta \P{\Omega^2}{\theta} = 
\frac{1}{\rho^2} \left(\P{\rho}{\theta}\P{p}{r} - \P{\rho}{r}\P{p}{\theta} \right).
\label{Eq:baroclinicity}
\end{eqnarray}
Using cylindrical coordinate, this equation can be  rewritten as
\begin{eqnarray}
 R \P{\Omega^2}{z} = \frac{1}{\rho^2 } \left( 
\P{\rho}{R}  \P{p}{z} - \P{\rho}{z} \P{p}{R} \right).
\end{eqnarray}
Therefore, we can solve the distribution of $\Omega$ by  
imposing a boundary condition on $\Omega$ at the equatorial 
surface and integrating these equations directly.
In this paper, we impose the following boundary condition on $\Omega$ at the 
equatorial surface:
\begin{eqnarray}
 \Omega^2(r, \pi/2) = \frac{j_0^2}{(1 + r^2 / A^2)^2},
\label{Eq:boundary}
\end{eqnarray}
where $j_0$, $A$ are constants. This is a $j$-constant rotation 
law (\citealt{Eriguchi_Muller_1985}). 

The physical quantities are transformed to dimensionless ones for the actual numerical 
computations using the central density $\rho_{c}$ and the equatorial radius $r_e$ as follows:
\begin{eqnarray}
 \hat{r} \equiv \frac{r}{r_e},
\end{eqnarray} 
\begin{eqnarray}
 \hat{\rho} \equiv \frac{\rho}{\rho_{c}},
\end{eqnarray}
\begin{eqnarray}
 \hat{\phi}  \equiv \frac{\phi}{4 \pi G r_e^2 \rho_{c}},
\end{eqnarray}
\begin{eqnarray}
 \hat{\Omega} \equiv \frac{\Omega}{\sqrt{4 \pi G \rho_{c}}},
\end{eqnarray}
\begin{eqnarray}
 \hat{K} \equiv \frac{K}{4\pi G r_e^2} \rho_c^{1 - 1/N},
\end{eqnarray}
where hat $\hat{}$ denotes a dimensionless quantity. 
Note that the dimensionless value of the central density is unity 
in this form since
\begin{eqnarray}
 \hat{\rho}_c = \frac{\rho_c}{\rho_c}= 1.
\label{Eq:rho_c}
\end{eqnarray}
The dimensionless pressure is obtained by the dimensionless form 
of the equation of state as
\begin{eqnarray}
 \hat{p} = \hat{K}_0 \left\{ 1 + \hat{\epsilon} 
\left( \frac{\sin^2 \theta}{\hat{a}_0^2} + \frac{\cos^2 \theta}{\hat{b}_0^2} \right) \hat{r}^m \right\} \hat{\rho}^{1 + 1/N},
 \label{Eq:p_k_rho2}
\end{eqnarray}
where $\hat{\epsilon}$, $\hat{a}_0$ and $\hat{b}_0$ are dimensionless parameters. 
Thus, the central pressure $\hat{p}_c$ can be obtained as
\begin{eqnarray}
 \hat{p}_c = \hat{K}_0 \hat{\rho}_c^{1 + 1/N}.
 \label{Eq:p_c}
\end{eqnarray}
The value of $\hat{K}_0$ is determined to ensure a unity distance 
from the centre to the equatorial surface of the star 
(\citealt{Tomimura_Eriguchi_2005}).
All equations are also transformed to be dimensionless.
For instance, the profile of the angular velocity (equation \ref{Eq:boundary}) is transformed as
\begin{eqnarray}
 \hat{\Omega}^2(\hat{r}, \pi/2) = \frac{\hat{j}_0^2}{ (1 + \hat{r}^2 / \hat{A}^2)^2},
  \label{Eq:boundary2}
\end{eqnarray}
where $\hat{j}_0$ and $\hat{A}$ are dimensionless parameters.
$\hat{A}$ denotes the degree of the differential rotation. 

 In summary, the  dimensionless forms of 
 equations (\ref{Eq:eular_r}), (\ref{Eq:Poisson_int}) and  (\ref{Eq:baroclinicity}) 
 have been defined with the dimensionless equation of state (equation \ref{Eq:p_k_rho2})
 for the actual numerical calculation. 

\subsection{Numerical scheme}

Next, the numerical scheme of the new method are described briefly.
The details of the numerical scheme is described in Appendix A.
The numerical scheme adopts a self-consistent field iteration scheme 
(\citealt{Ostriker_Mark_1968}). The gravitational field and all 
physical quantities are calculated iteratively in the self-consistent field scheme.
In particular, the numerical schemed of the new method is based  on 
 Hachisu's Self-Consistent Field (HSCF) scheme (\citealt{Hachisu_1986a,Hachisu_1986b}),
which is a well-established numerical method for 
calculating rapidly rotating barotropic equilibrium states.
However, the HSCF scheme cannot calculate rotating baroclinic stars 
because it uses the first integral of the Euler equation (equation \ref{Eq:rot_phi}).
In contrast, the numerical scheme of the new method uses the differential forms of the Euler equations
(equations \ref{Eq:eular_r}, \ref{Eq:baroclinicity}) and can calculate
rotating baroclinic stars self-consistently.

The new method calculate these equations explicitly (Appendix A), 
while the previous studies for rotating baroclinic stars 
(\citealt{Uryu_Eriguchi_1994, Uryu_Eriguchi_1995};
\citealt{Espinosa_Rieutord_2007, Espinosa_Rieutord_2013})
adopted implicit calculations (e.g. Newton Raphson method) 
or spectral-type method to solve these equations numerically. Since the 
implementation of the explicit method is much easier than 
those of the implicit method and spectral-type method,
the implementation of the new scheme is much easier than
those of the previous works. We can calculate 
rotating baroclinic stars easily and stably by using the new scheme.

Following the previous works,
the axis ratio is fixed during iteration cycles in the numerical scheme
(\citealt{Eriguchi_Muller_1985}; \citealt{Hachisu_1986a,Hachisu_1986b}). 
The axis ratio $q$ is defined as
\begin{eqnarray}
 q \equiv \frac{r_{\rm pol.}}{r_{\rm eq.}},
\end{eqnarray}
where $r_{\rm eq.}$ and $r_{\rm pol.}$ are 
the equatorial radius and the polar radius of the star, respectively.
Extremely rapidly rotating stars can be obtained stably 
by fixing the axis ratio $q$ (\citealt{Eriguchi_Muller_1985}; \citealt{Hachisu_1986a}).


\section{Numerical results}

Numerical results obtained by our novel method are displayed in this section.
The  accuracy of the method is verified in Section 3.1.1.
In Section 3.1.2, numerical results are compared with results obtained via 
previous method to confirm the numerical results.
Two types of rotating baroclinic sequences are investigated by fixing the functional forms 
of the baroclinic equation of state with
these new solutions displayed and discussed in Section 3.2.

\subsection{Accuracy verification}

\subsubsection{Virial test}

Numerical accuracy and convergence checking are important when investigating
a new numerical method. To check the numerical accuracy,
a relative value of the virial relation was computed as follows:
\begin{eqnarray}
 {\rm VC} \equiv \frac{|2T + 3\Pi + W|}{|W|},
\end{eqnarray}
where $T$, $\Pi$ and $W$ are, respectively, 
 the rotational energy, the total pressure of the star 
and the gravitational energy with
\begin{eqnarray}
 T = \frac{1}{2}\int \rho (r\sin \theta)^2 dV,
\end{eqnarray}
\begin{eqnarray}
 W = \frac{1}{2}\int \phi \rho dV,
\end{eqnarray}
\begin{eqnarray}
 \Pi = \int p dV.
\end{eqnarray}
A numerical domain is defined as $0 \leq \hat{r} \leq 2 $ in the
radial direction and $0 \leq \theta \leq \pi/2$ in the angular direction. 
The number of the grid points in the $r$ 
direction within $\hat{r} \in [0,1]$
and that in the $\theta$ direction within $\theta \in [0, \pi/2]$ are defined 
 as $N_r + 1$  and $N_\theta$, respectively.  
The numerical domains are discretized into mesh points with 
equal intervals $\Delta\hat{r}$ and $\Delta \theta$ (equations \ref{Eq:app1} and \ref{Eq:app2}).
Therefore, the total mesh number is  $2 N_r \times N_\theta + 1$ in the computations. 
$N_\theta = 129$ was fixed during the test computations and
the value of $N_r$ was changed. Both barotropic ($\hat{\epsilon} = 0$)
and baroclinic ($\hat{\epsilon} = -0.05$) solutions were calculated 
with $q = 0.9$, $\hat{a}_0 = \hat{b}_0 = 1.0$ and $N=1.5$.
The same model was also calculated using previous codes based on HSCF scheme
(\citealt{Fujisawa_Yoshida_Eriguchi_2012}; \citealt{Fujisawa_Eriguchi_2013}; \citealt{Fujisawa_2015a}) for comparison.

\begin{figure*}
\includegraphics[width=10cm]{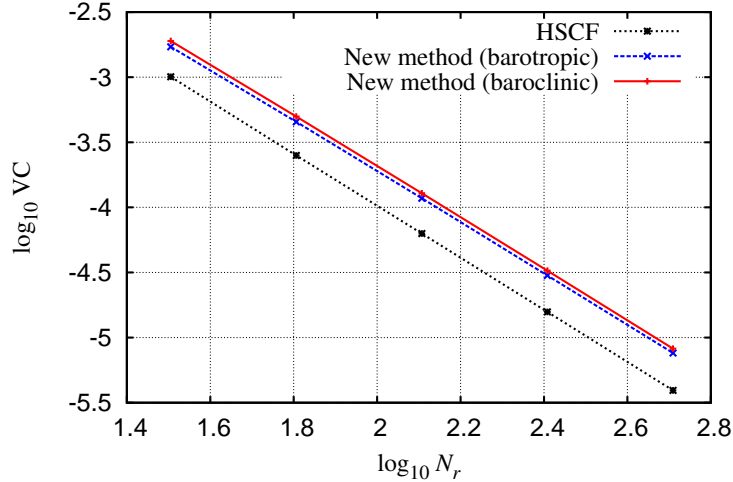}
 \caption{The values of VC versus the number of the grid points in the $r$ direction ($N_r$).
 Each line denotes rotating barotropic solutions by the new method (dashed line), rotating baroclinic solutions 
 by the new method (solid line) and barotropic solutions by the HSCF scheme (dotted line). }
 \label{Fig:VC}
\end{figure*}

Fig. \ref{Fig:VC} displays the convergence of the numerical results obtained 
via  each numerical method. The lines denote
the rotating baroclinic solutions using the new method (solid line), 
the rotating barotropic solutions using the new method (dashed line) and  
the rotating barotropic solutions using the HSCF scheme (dotted line). 
As seen in Fig. \ref{Fig:VC}, the values of VC become smaller as the 
grid number in the $r$ direction increases. 
Both the barotropic (dotted line) and baroclinic (dashed lined) solutions 
obtained by the new method converged well.
Although the values of the VC obtained by the new method are slightly 
larger than that obtained by HSCF scheme, they are sufficiently small 
and the solutions converged well (see \citealt{Hachisu_1986a, Hachisu_1986b}). 
Therefore, the numerical results by the new method are accurate and 
the numerical accuracy can be considered verified.
In actual numerical calculations, which are tabulated 
and displayed in this paper, $N_r = 512$ and $N_\theta = 257$ are used.

\subsubsection{Comparison with HSCF scheme}

\begin{figure*}
   \includegraphics[width=8cm]{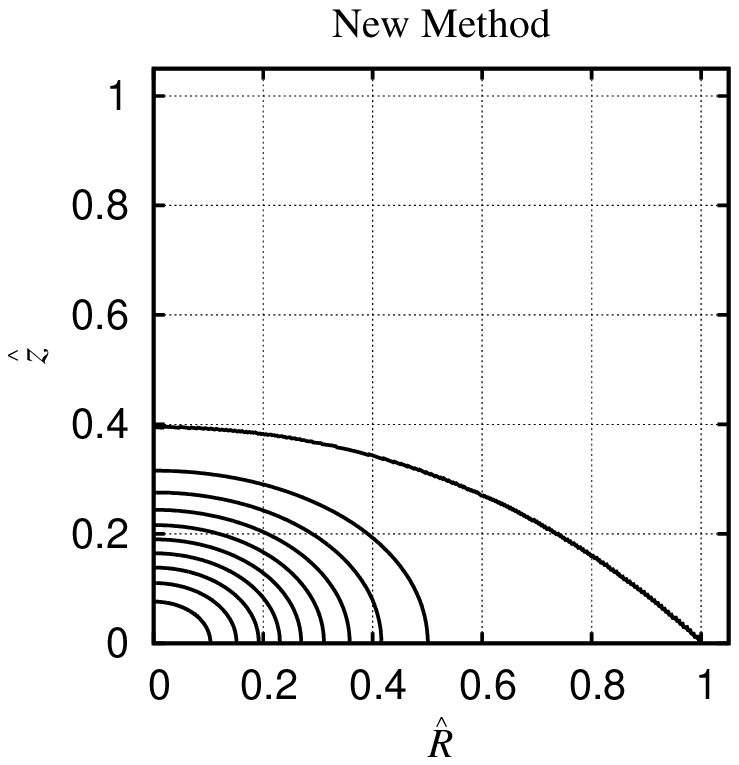}
   \includegraphics[width=8cm]{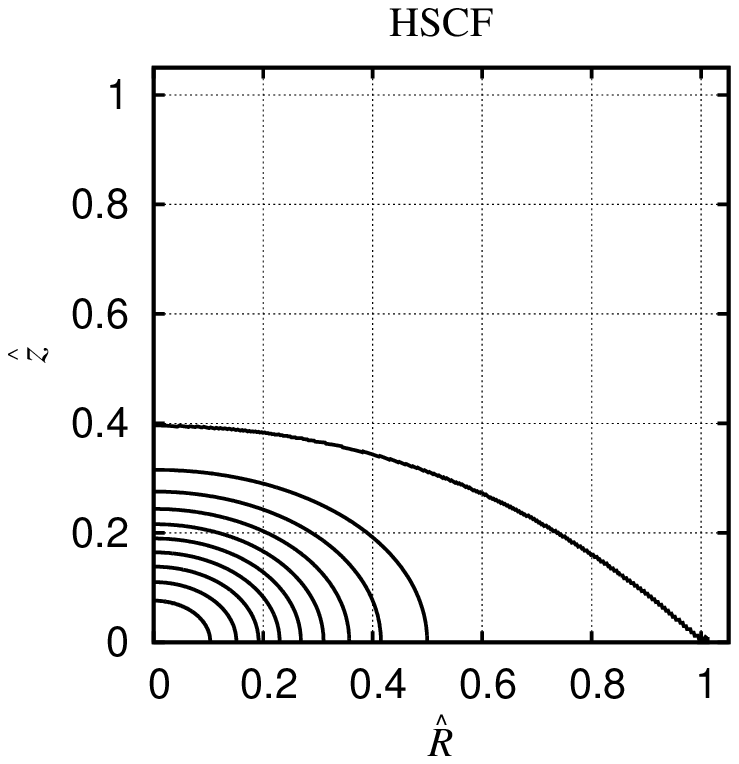}
   \caption{The isopycnic surfaces of the critical rotation models ($q = 0.395$)
   calculated by the new method (left-hand panel) and the HSCF scheme (right-hand panel). 
   The outermost curves denote the stellar surfaces.
   The difference between the two adjacent contours is $1/10$ of the
   maximum value of $\hat{\rho}$.}
   \label{Fig:New_HSCF}
  \end{figure*}

Next, the solutions obtained by the new numerical method were compared 
with those obtained by HSCF scheme to confirm the results. 
The baroclinic parameter was set as $\hat{\epsilon} = 0$ 
to  calculate the rotating barotropic equilibrium states by the new method.
The same parameters ($N = 1.5$ and $\hat{A} = 0.9$) were used  
and seven models of rotating barotropic equilibrium states were calculated via 
both the new method and HSCF scheme. 
The seven solutions by each code are tabulated in Table \ref{Tab:tab1}.
The left part of the table contains the numerical results obtained by
the new method and the right one contains those obtained by HSCF scheme.
The '$*$' in the table denotes the critical rotation model 
beyond which no rotating equilibrium state exists (\citealt{Eriguchi_Muller_1985}; 
\citealt{Hachisu_1986a}). The isopycnic surfaces of the critical models are 
displayed in Fig. \ref{Fig:New_HSCF}.

As seen in Table \ref{Tab:tab1}, the numerical results of the new method (left-hand side)
are almost the same as those of the HSCF scheme (right-hand side).
Although the values of $\hat{j}_0^2$ are slightly different 
the values of $\hat{K}_0$ and $T/|W|$ are almost identical. 
The critical rotation model was also obtained using the new method. 
Fig. \ref{Fig:New_HSCF} shows the isopycnic surfaces of the 
critical rotation models obtained by the new method (left-hand panel) 
and the HSCF scheme (right-hand panel). 
These isopycnic surfaces are almost identical as seen in the figures.
The solutions of the critical rotation models 
have cusp structure at the equatorial surface (see right-hand panel). 
The cusp structure is calculated correctly by the new method (left-hand panel).
Thus, the new method can calculate rapidly rotating barotropic stars
correctly and accurately. The reliability of the new method is thus confirmed 
by this comparison with HSCF scheme. 

  \begin{table*}
   \caption{Physical quantities obtained by applying the new numerical method (left-hand) and 
   the HSCF scheme (right-hand). The same parameters in both codes were used in both codes
   ($N=1.5$, $\hat{A} = 0.9$). The solutions with '*' denote 
   the critical rotation models beyond which no rotating equilibrium state exists.}
   
  \begin{tabular}{ccccccccc}
 \hline
  $q$ & $\hat{K}_0$   &$\hat{j}_0^2$ &  $T/|W|$ & VC \\
  \hline
&&&&  \hspace{-160pt} New numerical method \\
 \hline
  0.900 & 2.691E$-$2 & 1.853E$-$2 & 2.232E$-$2  & 7.608E$-$6 \\
  0.801 & 2.368E$-$2 & 3.617E$-$2 & 4.698E$-$2  & 7.712E$-$6 \\
  0.699 & 2.020E$-$2 & 5.255E$-$2 & 7.492E$-$2  & 7.925E$-$6 \\
  0.600 & 1.662E$-$2 & 6.553E$-$2 & 1.048E$-$1  & 8.262E$-$6 \\ 
  0.500 & 1.272E$-$2 & 7.234E$-$2 & 1.359E$-$1  & 9.087E$-$6 \\
  0.400 & 7.737E$-$2 & 5.616E$-$2 & 1.351E$-$1  & 1.460E$-$5 \\
  0.395* & 7.126E$-$2 & 5.107E$-$2 & 1.241E$-$1  & 1.658E$-$5 \\
  \hline
 \end{tabular}
 \begin{tabular}{ccccccccc}
 \hline
  $q$ & $\hat{K}_0$   &$\hat{j}_0^2$ &  $T/|W|$ & VC \\
\hline
&&&&  \hspace{-160pt} HSCF \\
 \hline
  0.900 & 2.691E$-$2 & 1.835E$-$2 & 2.210E$-$2  & 3.929E$-$6 \\
  0.801 & 2.368E$-$2 & 3.595E$-$2 & 4.668E$-$2  & 4.230E$-$6 \\
  0.699 & 2.021E$-$2 & 5.233E$-$2 & 7.458E$-$2  & 4.646E$-$6 \\
  0.600 & 1.661E$-$2 & 6.543E$-$2 & 1.047E$-$1  & 5.234E$-$6 \\
  0.500 & 1.271E$-$2 & 7.225E$-$2 & 1.357E$-$1  & 6.269E$-$6 \\
  0.400 & 7.702E$-$3 & 5.589E$-$2 & 1.345E$-$1  & 1.030E$-$5 \\
  0.395* & 7.109E$-$3 & 5.093E$-$2 & 1.238E$-$1  & 1.140E$-$5 \\
 \hline
  \end{tabular}
   \label{Tab:tab1}
  \end{table*}

\subsection{Rapidly rotating baroclinic equilibrium states}

Finally, the rapidly rotating baroclinic equilibrium states were calculated 
using the new method. Two types of rotating baroclinic 
models were developed: a spherical isentropic surfaces model and  
an oblate isentropic surfaces model. 
The solution sequences of these two models were calculated
systematically and the critical rotation solutions of these models were 
obtained via  the new method. The Bjerknes--Rosseland rules, which must be 
satisfied by rotating baroclinic stars (c.f. \citealt{Tassoul_1978}), were checked,
and all of the solutions were found to them.

\subsubsection{Spherical isentropic surfaces model ($\hat{a}_0 = \hat{b}_0$)}

\begin{figure*}

\includegraphics[width=5.4cm]{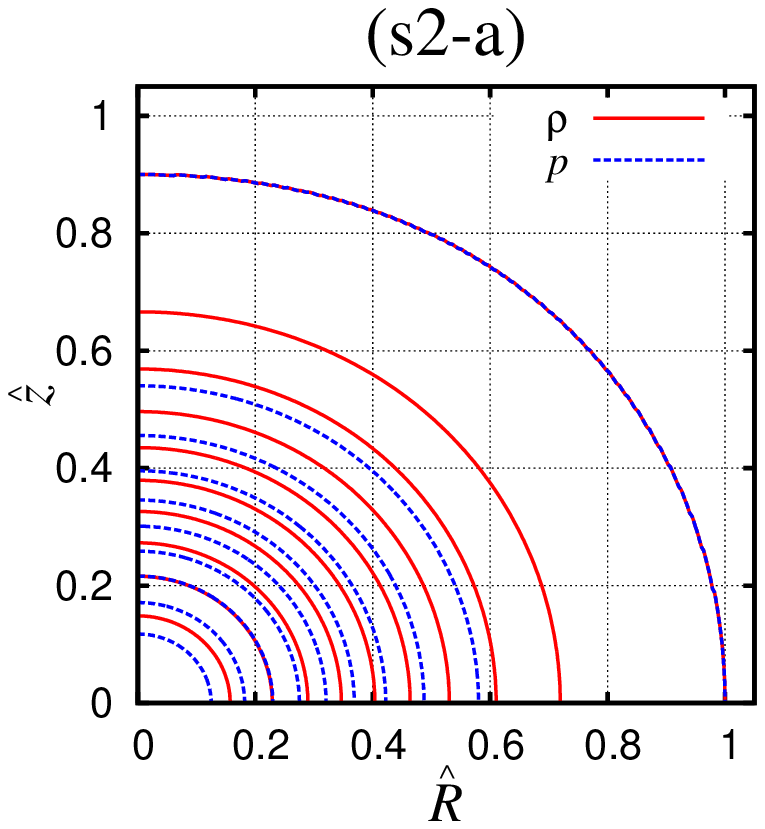}
\includegraphics[width=5.4cm]{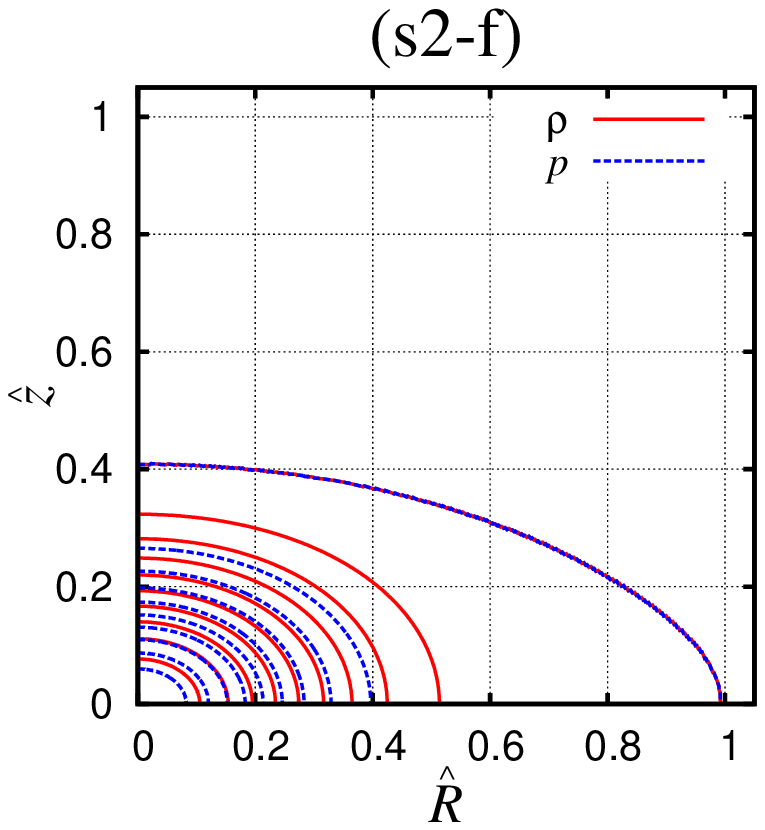}
\includegraphics[width=5.4cm]{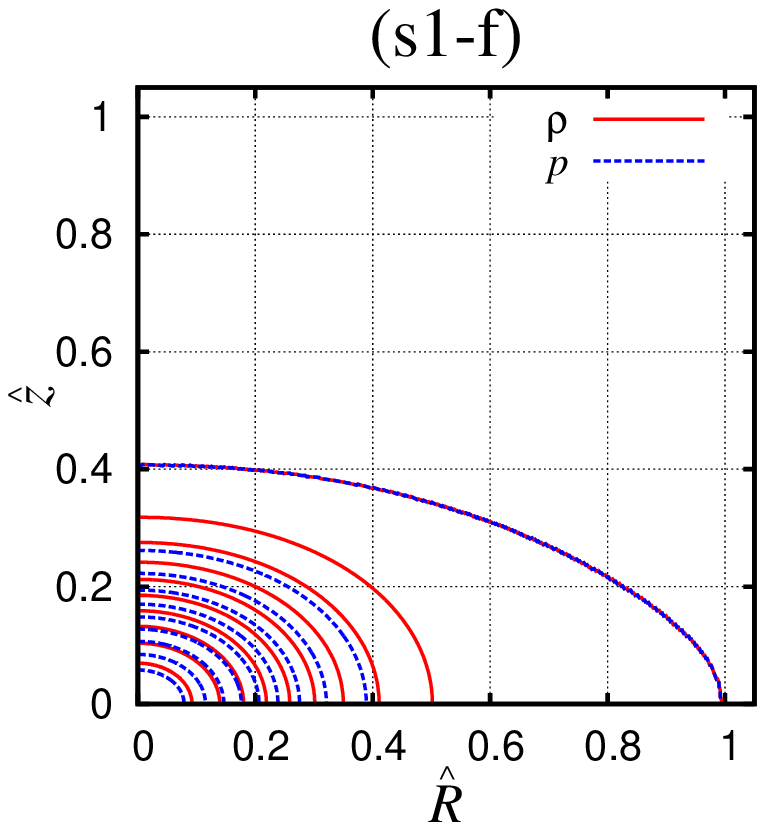}

\includegraphics[width=5.4cm]{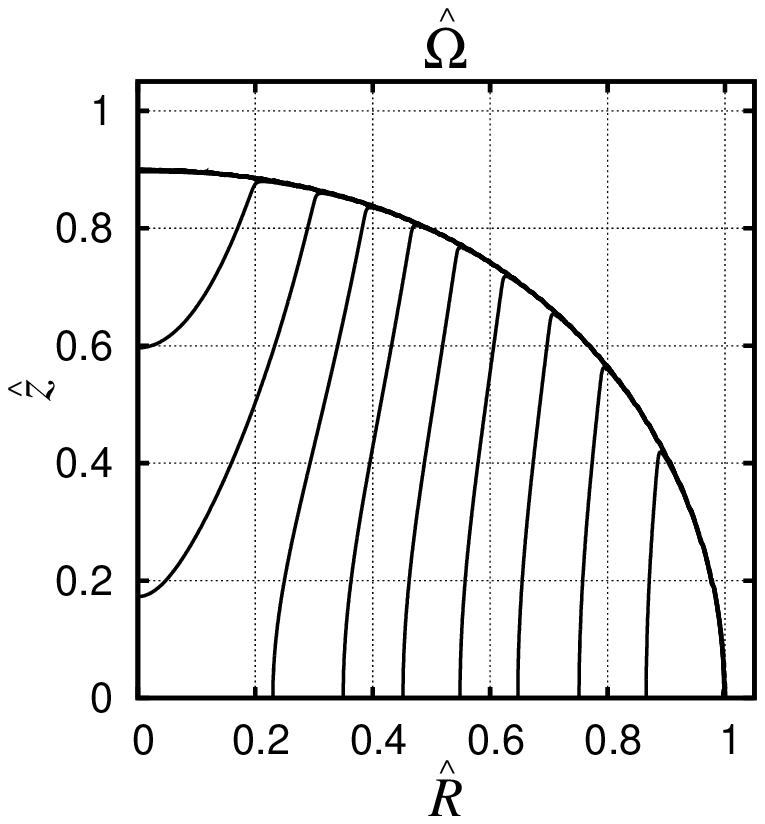}
\includegraphics[width=5.4cm]{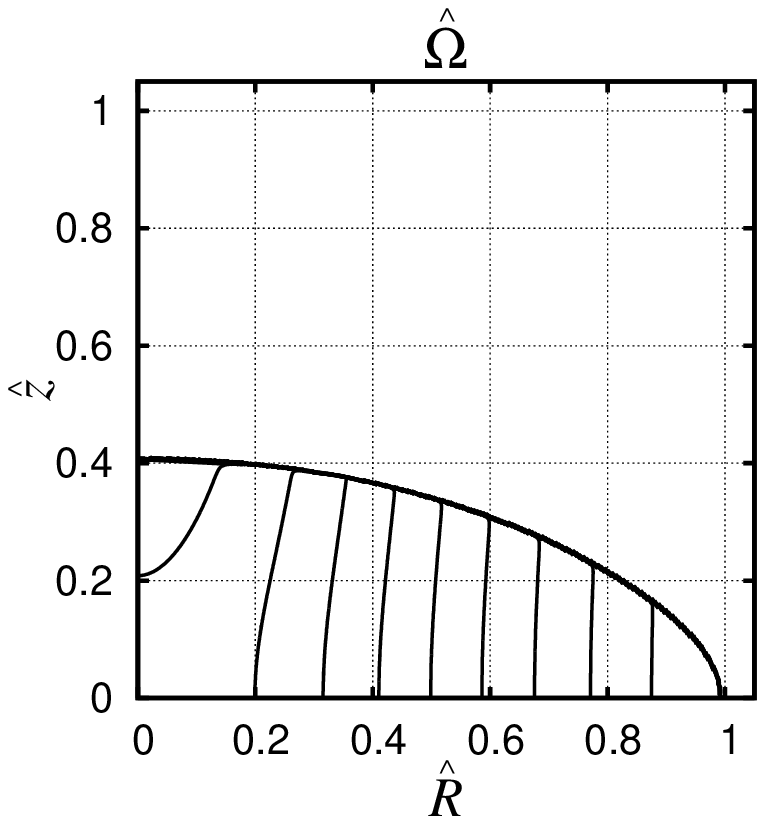}
\includegraphics[width=5.4cm]{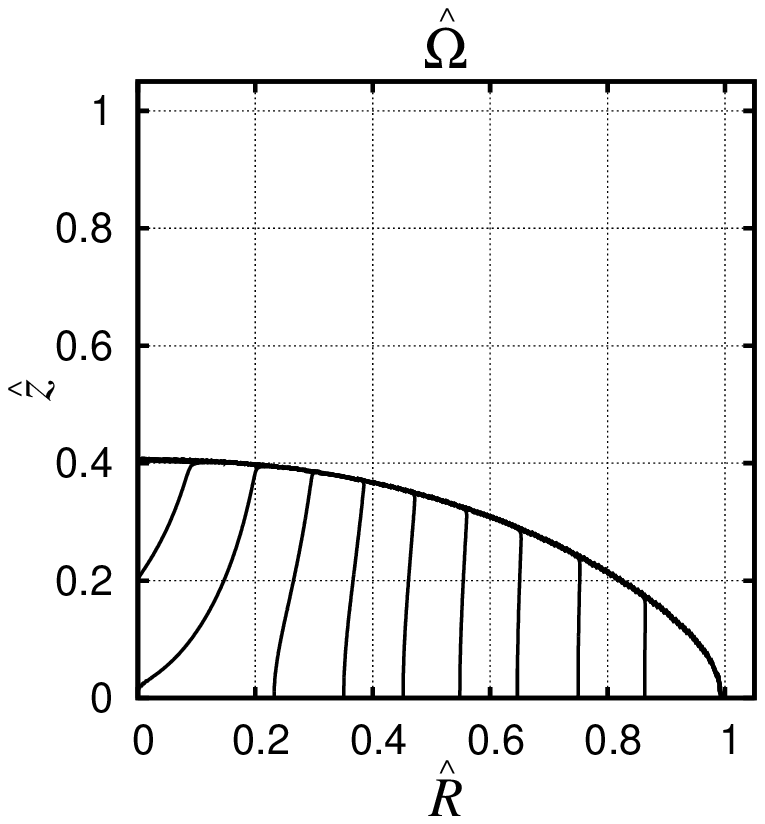}

\includegraphics[width=5.4cm]{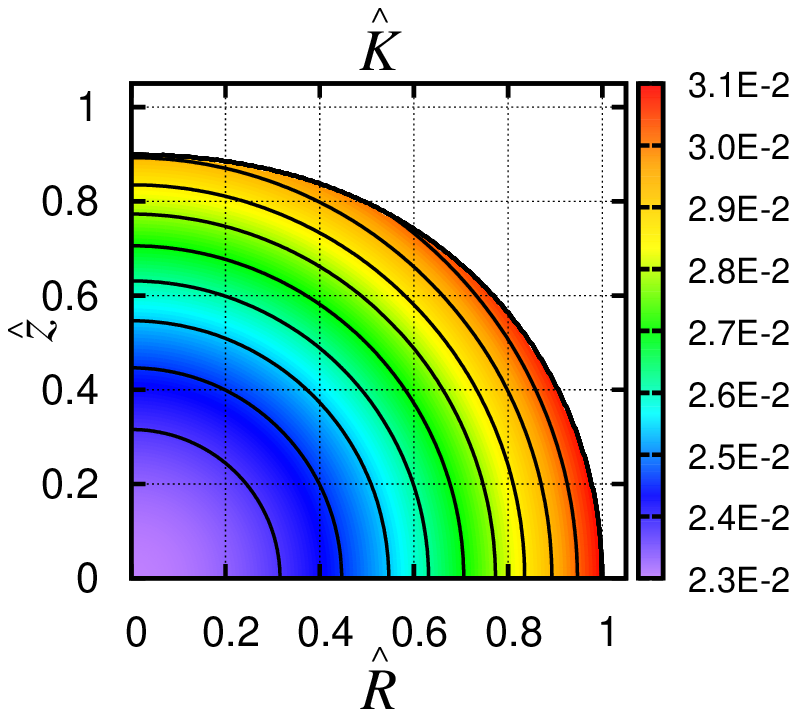}
\includegraphics[width=5.4cm]{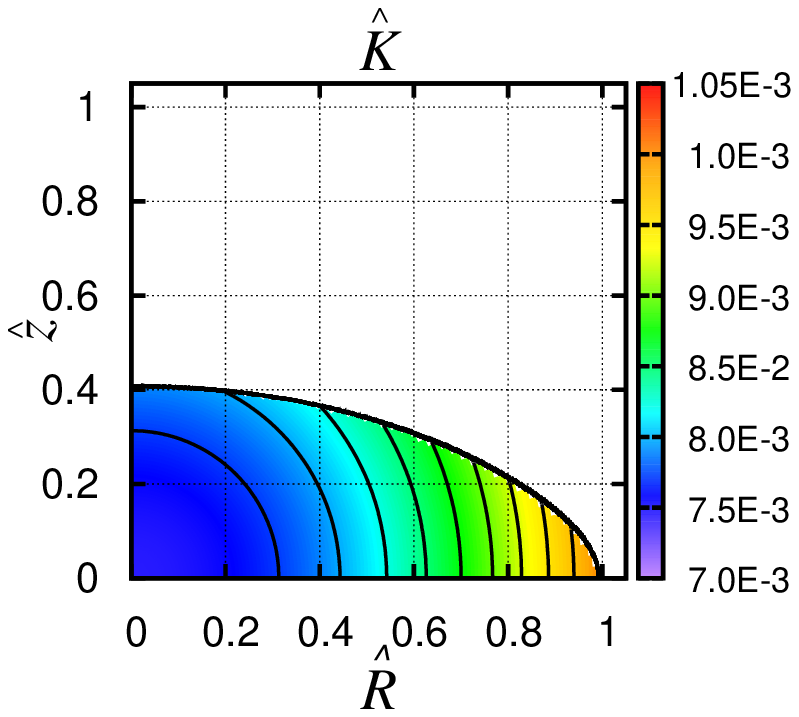}
\includegraphics[width=5.4cm]{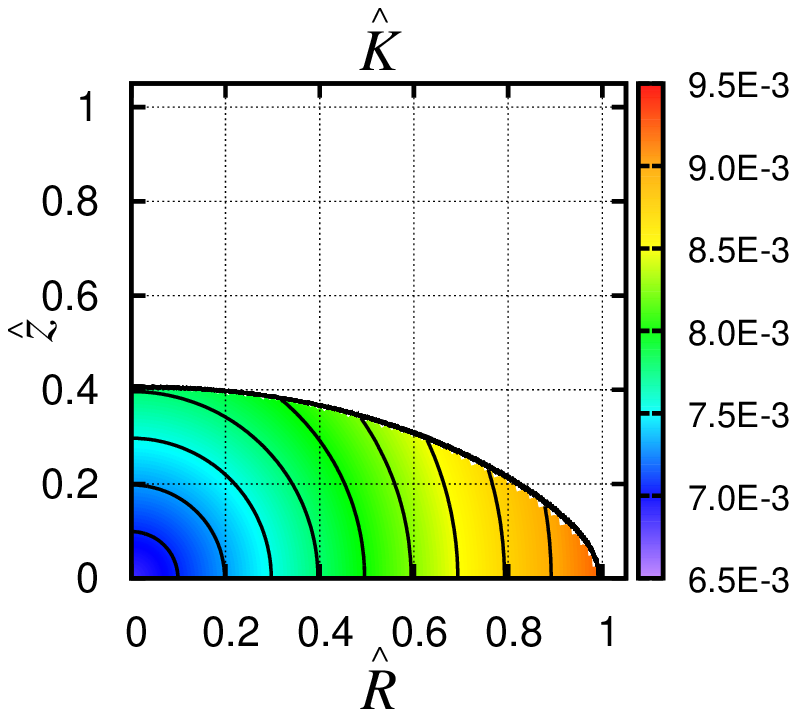}

\includegraphics[width=5.4cm]{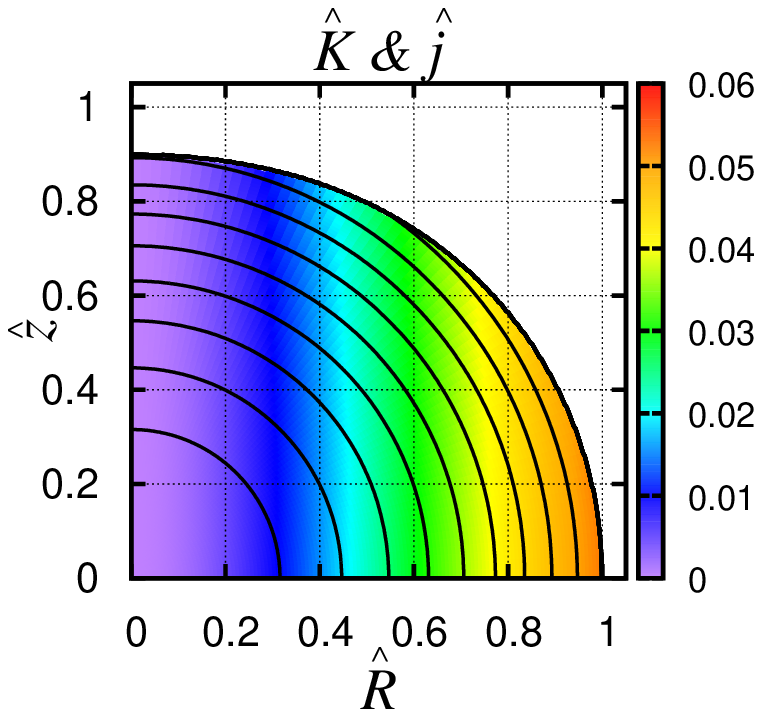}
\includegraphics[width=5.4cm]{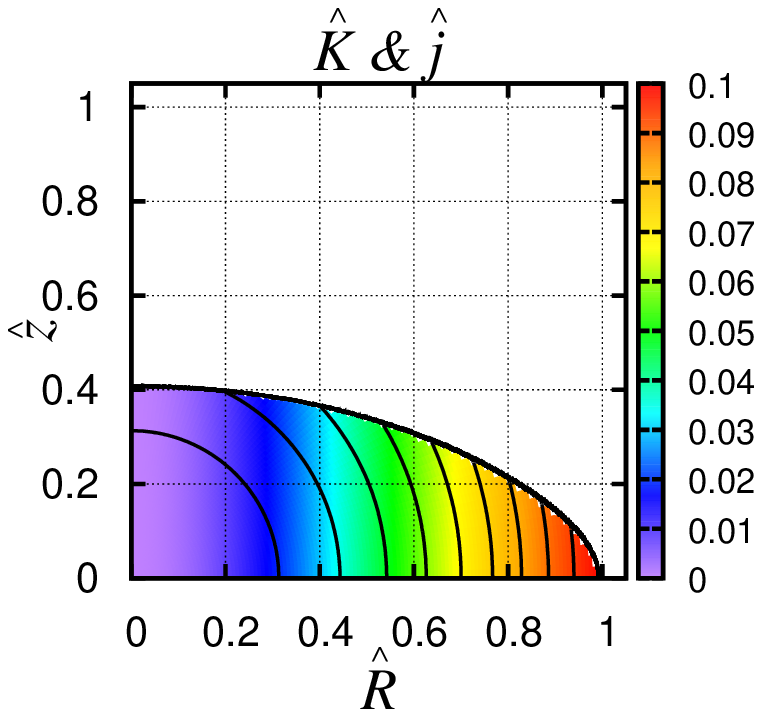}
\includegraphics[width=5.4cm]{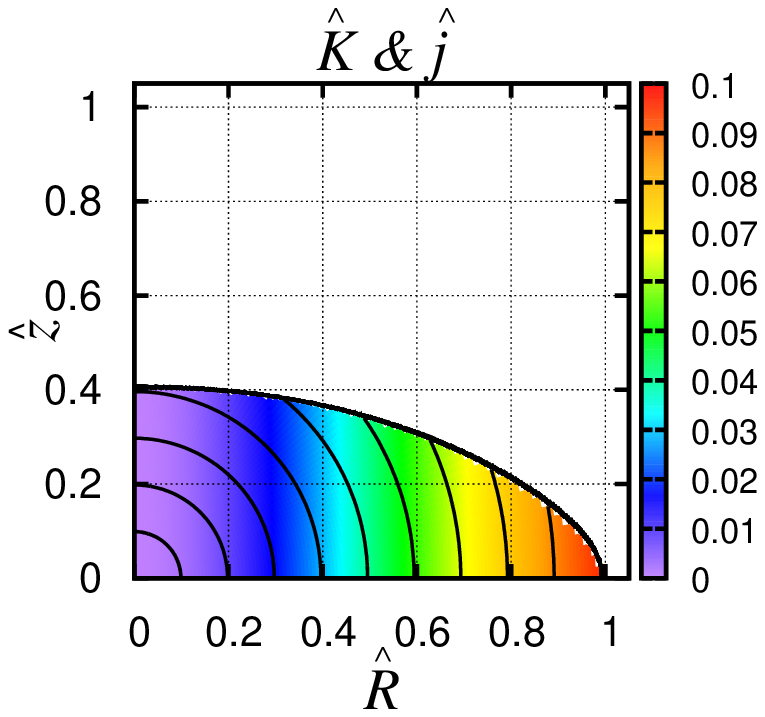}

\caption{Distributions of the physical quantities. 
 The three columns of figures display the distributions of solution (s2-a) (left-hand panels), 
 solution (s2-f) (central panels) and solution (s1-f) (right-hand panels).
 From top to bottom, the four rows of figures correspond to 
 the isopycnic surfaces (solid curves) and isobaric surfaces (dashed curves) (first row),
 the angular velocity isosurfaces (second row), 
 the $\hat{K}$ isosurfaces (contours) and the value of $\hat{K}$ (colour maps) (third row) and
 the $\hat{K}$ isosurfaces (contours) and the value of the dimensionless specific 
 angular momentum $\hat{j}$ (fourth row). The difference between the two adjacent contours is 1/10 of 
the difference between the maximum value and the minimum value.}
 \label{Fig:contours1}
\end{figure*}


\begin{table*}
 \caption{Physical quantities for the sequences with the spherical isentropic surfaces.
The solutions with '*' denote the critical rotation models.}
 \begin{tabular}{cccccccccccc}
 \hline
&$q$&  $\hat{\epsilon}$ & $\hat{a}_0$ & $\hat{b}_0$ & $m$ &  $\hat{K}_0$   & $\hat{j}_0^2$ &  $T/|W|$ & $\Pi / |W|$ & VC \\
\hline
(s2-a) & 0.900 & 0.35 & 1.0 & 1.0 & 2.0 & 2.300E$-$2 & 1.390E$-$2 & 1.910E$-2$ & 3.206E$-1$ & 8.447E$-6$ \\
(s2-b)& 0.801 & 0.35 & 1.0 & 1.0 & 2.0 & 1.980E$-$2 & 2.611E$-$2 & 3.940E$-2$ & 3.071E$-1$ & 8.817E$-6$ \\
(s2-c)& 0.699 & 0.35 & 1.0 & 1.0 & 2.0 & 1.776E$-$2 & 4.136E$-$2 & 6.579E$-2$ & 2.895E$-1$ & 8.816E$-6$ \\
(s2-d)& 0.599 & 0.35 & 1.0 & 1.0 & 2.0 & 1.472E$-$2 & 5.274E$-$2 & 9.342E$-2$ & 2.711E$-1$ & 9.285E$-6$ \\
(s2-e)& 0.500 & 0.35 & 1.0 & 1.0 & 2.0 & 1.133E$-$2 & 5.887E$-$2 & 1.212E$-2$ & 2.526E$-1$ & 1.046E$-5$ \\
(s2-f)& 0.408* & 0.35 & 1.0 & 1.0 & 2.0 & 7.462E$-$3 & 5.009E$-$2 & 1.238E$-2$ & 2.508E$-1$ & 1.559E$-5$ \\
\hline
(s1-a) & 0.900 & 0.35 & 1.0 & 1.0 & 1.0 & 2.133E$-$2 & 1.369E$-$2 & 1.967E$-2$ & 3.202E$-1$ & 8.354E$-6$ \\
(s1-b) & 0.801 & 0.35 & 1.0 & 1.0 & 1.0 & 1.899E$-$2 & 2.718E$-$2 & 4.174E$-2$ & 3.055E$-1$ & 8.508E$-6$ \\
(s1-c) & 0.699 & 0.35 & 1.0 & 1.0 & 1.0 & 1.639E$-$2 & 4.009E$-$2 & 6.706E$-2$ & 2.886E$-1$ & 8.757E$-6$ \\
(s1-d) & 0.599 & 0.35 & 1.0 & 1.0 & 1.0 & 1.359E$-$2 & 5.062E$-$2 & 9.452E$-2$ & 2.703E$-1$ & 9.239E$-6$ \\
(s1-e) & 0.500 & 0.35 & 1.0 & 1.0 & 1.0 & 1.047E$-$2 & 5.601E$-$2 & 1.219E$-1$ & 2.520E$-1$ & 1.047E$-5$ \\
(s1-f) & 0.406* & 0.35 & 1.0 & 1.0 & 1.0 & 6.833E$-$3 & 4.662E$-$2 & 1.225E$-1$ & 2.517E$-1$ & 1.607E$-5$ \\
\hline
 \end{tabular}
 \label{Tab:tab2}
\end{table*}

First, the rotating baroclinic stars with spherical isentropic 
surfaces were calculated as one of the simplest baroclinic models. 
The parameters of the equation of the state were fixed as $\hat{a}_0 = \hat{b}_0 = 1.0$ to
make a spherical isentropic surfaces. 
Two solution sequences (solutions (s1-) and (s2-)) with different $m$ were calculated 
by changing the value of $q$,  and ten solutions (solutions (s1-a)--(s1-e) and (s2-a)--(s2-e)) 
and two baroclinic critical rotation solutions (solutions (s1-f) and (s2-f)) were obtained. 
The physical quantities of the numerical results are tabulated in Table \ref{Tab:tab2}.  
The solutions with '*' in Table \ref{Tab:tab2} denote the critical rotation models.
The structures of the solutions and distributions of the physical quantities 
are displayed in Fig. \ref{Fig:contours1} with
solution (s2-a) in the left-hand column, (s2-f) in the central column 
and (s1-f) in the right-hand column.  
From top to bottom, the four rows in Fig. \ref{Fig:contours1} 
correspond to the isopycnic surfaces (solid curves) 
and isobaric surfaces (dashed curves) in the first row,
 the angular velocity isosurfaces in the second row, 
 the $\hat{K}$ isosurfaces (contours) and the value of $\hat{K}$ (colour maps) in the third row and
 the $\hat{K}$ isosurfaces (contours) and the value of the dimensionless specific 
 angular momentum $\hat{j}$ in the fourth row. 
 The isopycnic surfaces and isopycnic surfaces of solution (s2-e) are also displayed in 
 the left-hand panel of Fig. \ref{Fig:iso_surfaces}
 to check the baroclinicity easily. Note that Fig. \ref{Fig:iso_surfaces} shows the 
 arbitrary seven isosurfaces and does not indicate the distributions of density and pressure.

As seen in the isobaric surfaces (solid curves) and 
the isopycnic surfaces in the left-hand panel of Fig. \ref{Fig:iso_surfaces}, 
the isobaric surfaces are always more oblate 
than the isopycnic surfaces.
This tendency is also found in the first row in Fig. \ref{Fig:contours1}
(particularly solution s2-f). 
This baroclinicity comes from the entropy distributions inside the star.
The third row shows the $\hat{K}$ isosurfaces (contours) and the 
values of $\hat{K}$ (colour maps). 
The $\hat{K}$ isosurfaces are spherical due to the functional form of equation (\ref{Eq:p_k_rho2}).
If we assume that the baroclinic ideal gas stars, 
these solutions have spherical isentropic surfaces
because the specific entropy $s$ is proportional to $\hat{K}$ 
and the $\hat{K}$ isosurfaces denote the isentropic surfaces.
The distributions of $\hat{K}$ in Fig. \ref{Fig:contours1} indicate 
that the specific entropy increases outward and the equatorial specific entropy is higher than the polar specific entropy.
Therefore,  the surface temperature is colder at the poles than at the equator.

The second row shows the angular velocity ($\hat{\Omega}$) distributions.
The equatorial angular velocity decreases outward
because the profile of the angular velocity on the equatorial plane is fixed as equation (\ref{Eq:p_k_rho2}).
On the other hand, the angular velocity on the rotational axis increases from 
the centre to the surface ($\partial \hat{\Omega} / \partial \hat{z} > 0$).
Moreover, the values of $\partial \hat{\Omega} / \partial \hat{z}$ are always positive 
inside the star. The angular velocity distribution of a barotropic star 
is always purely cylindrical (Fig. \ref{Fig:New_HSCF}), but baroclinicity causes the 
 angular velocity distribution of a baroclinic star to not be cylindrical.
 The distribution is determined by the inclination angle of the isobaric and the isopycnic surfaces 
of the rotating star (equation \ref{Eq:baroclinicity}).
According to the Bjerknes--Rosseland rules, 
the value of $\partial \Omega / \partial z$ is always positive if the
isobaric surfaces are more oblate than the isopycnic surfaces 
and the surface temperature is colder at the poles than at the equator (\citealt{Tassoul_1978}).
All of these rotating baroclinic stars satisfy this rule correctly.
The fourth row shows the $\hat{K}$ isosurfaces (contours) and 
the specific angular momentum $\hat{j} = (r\sin\theta)^2 \Omega$ 
distributions (colour map). As seen in these panels, 
the specific angular momentum on each $\hat{K}$ isosurface ($s$ = constant surface) 
increases as we move from poles to the equator. 

The axis ratios $q$ of both critical models are slightly larger than that of 
the rotating barotropic  model (Table \ref{Tab:tab1}). 
The energy ratios $T/|W|$ of the critical rotating baroclinic stars are slightly
smaller than that of barotropic one. The central and right-hand panels in Fig. \ref{Fig:contours1}
shows the critical rotation model with $m = 2$ (central panels) and $m=1$ (right-hand panels).
Although the isentropic surfaces of $m = 1$ (s1-f) are more concentrated than those 
of $m = 2$ (s2-f), the angular velocity distributions are similar. 
The parameter $m$ does change the distributions of the angular velocity very much.
Therefore, the shape of the isentropic surfaces is more important for determining the 
distribution of the angular velocity.

Summarizing the above, the rotating baroclinic stars have spherical isentropic 
surfaces and the isobaric surfaces are more oblate than the isopycnic surfaces 
due to the entropy distribution. Such baroclinicity changes the angular velocity distributions 
and the value of $\partial \Omega / \partial z$ is always positive inside the star.
This condition itself is one of the  Bjerknes--Rosseland rule itself (\citealt{Tassoul_1978}).  
Therefore, all of these rotating baroclinic stars satisfy the Bjerknes--Rosseland rules.

\subsubsection{Oblate isentropic surfaces models ($\hat{a}_0 > \hat{b}_0 $)}


\begin{figure*}
\includegraphics[width=5.4cm]{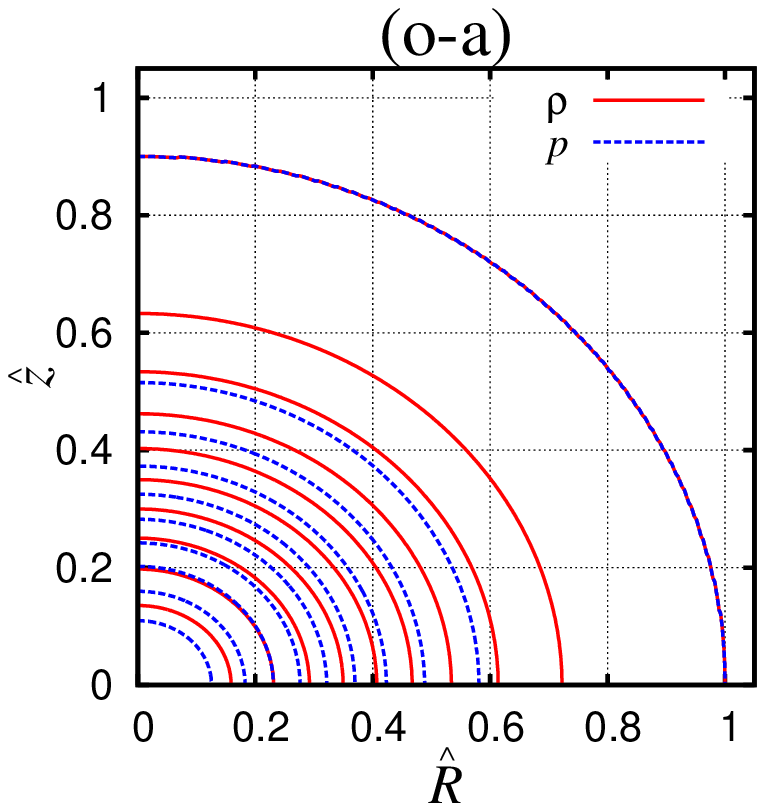}
\includegraphics[width=5.4cm]{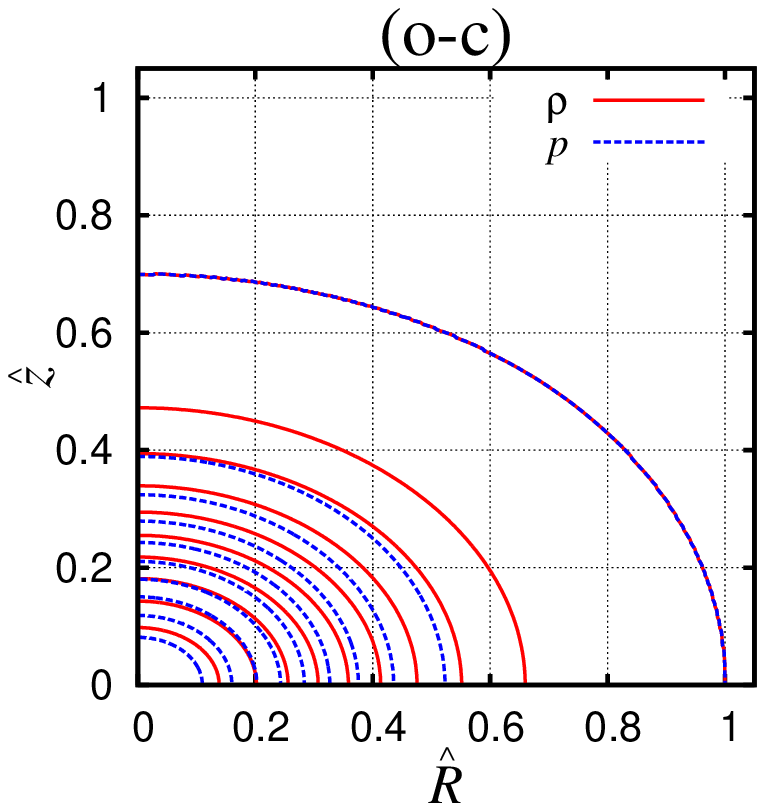}
\includegraphics[width=5.4cm]{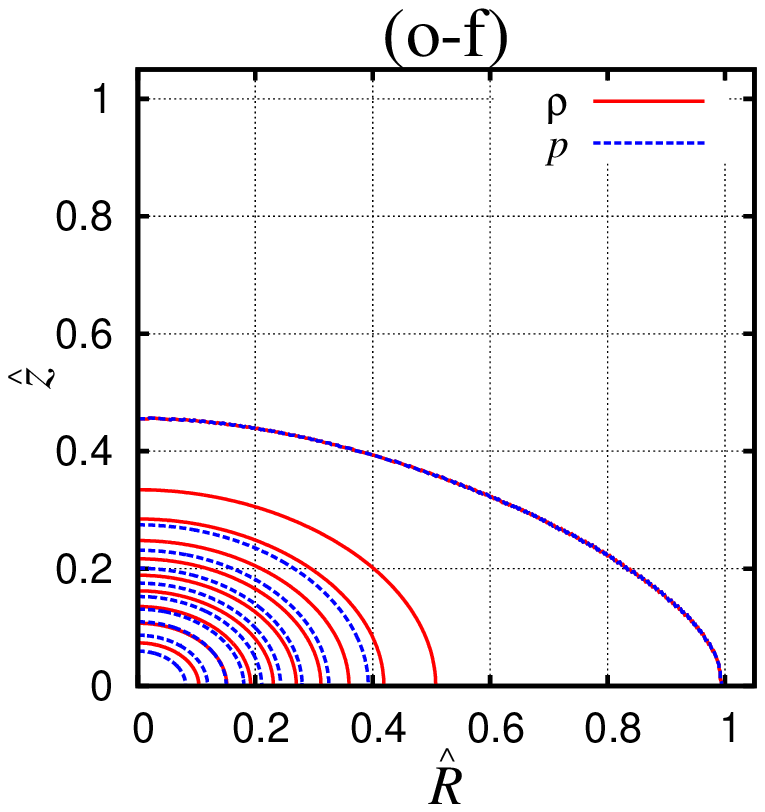}

\includegraphics[width=5.4cm]{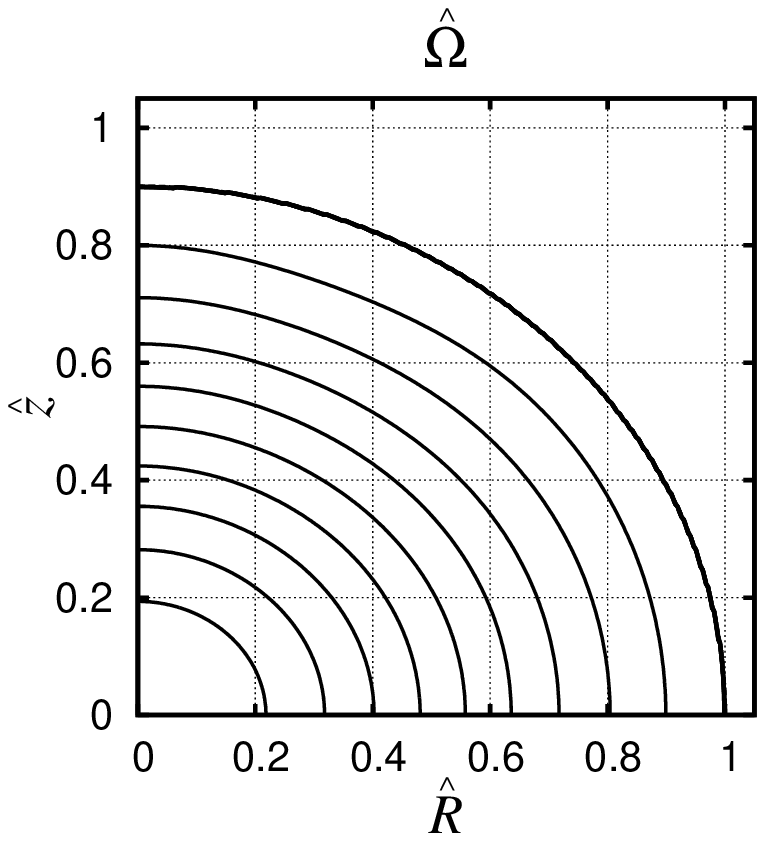}
\includegraphics[width=5.4cm]{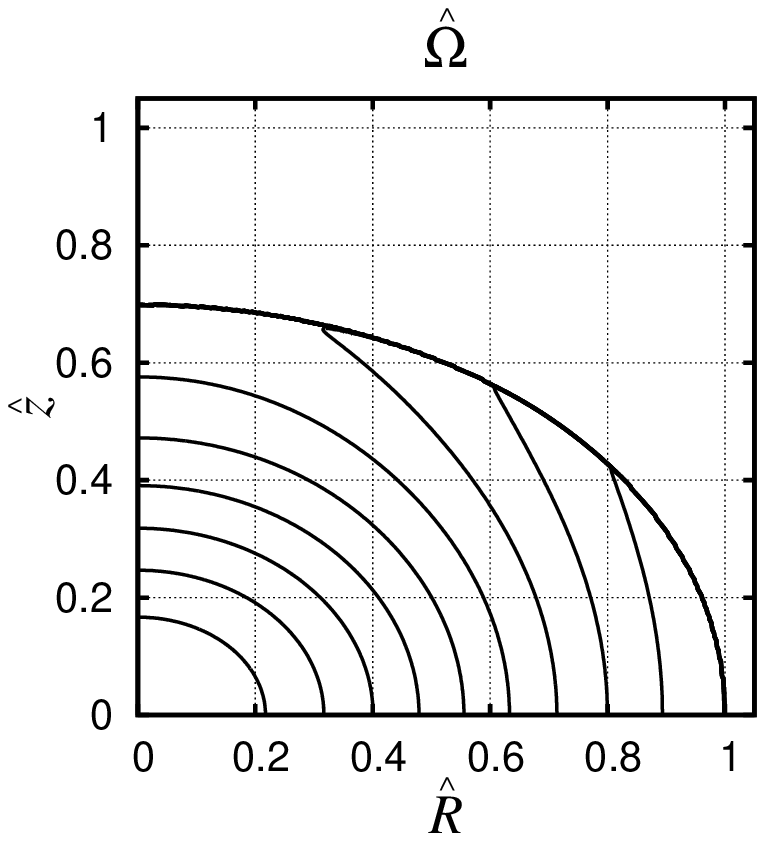}
\includegraphics[width=5.4cm]{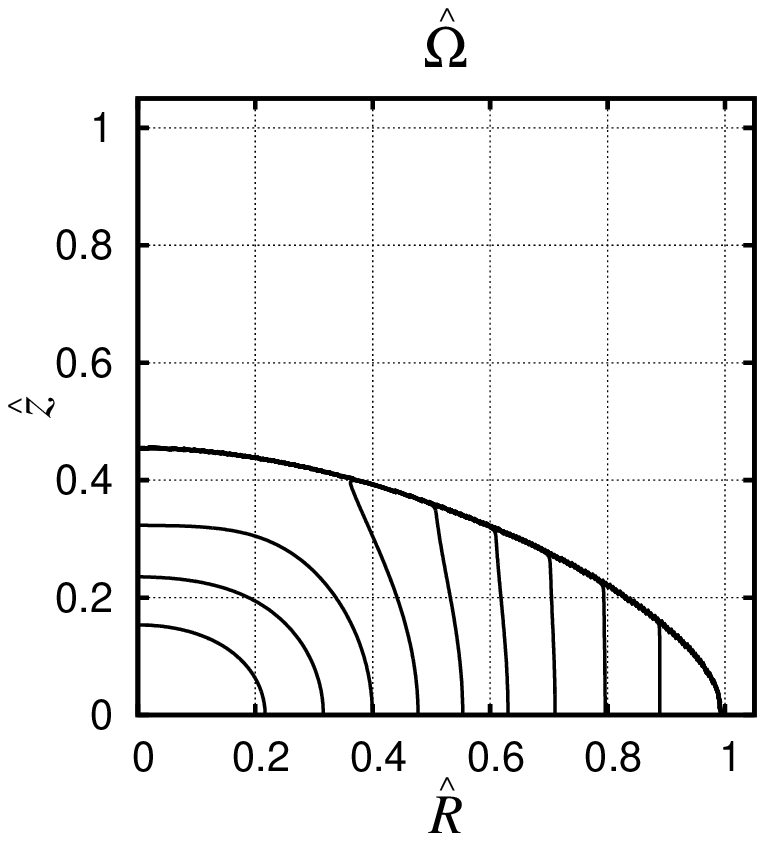}

\includegraphics[width=5.4cm]{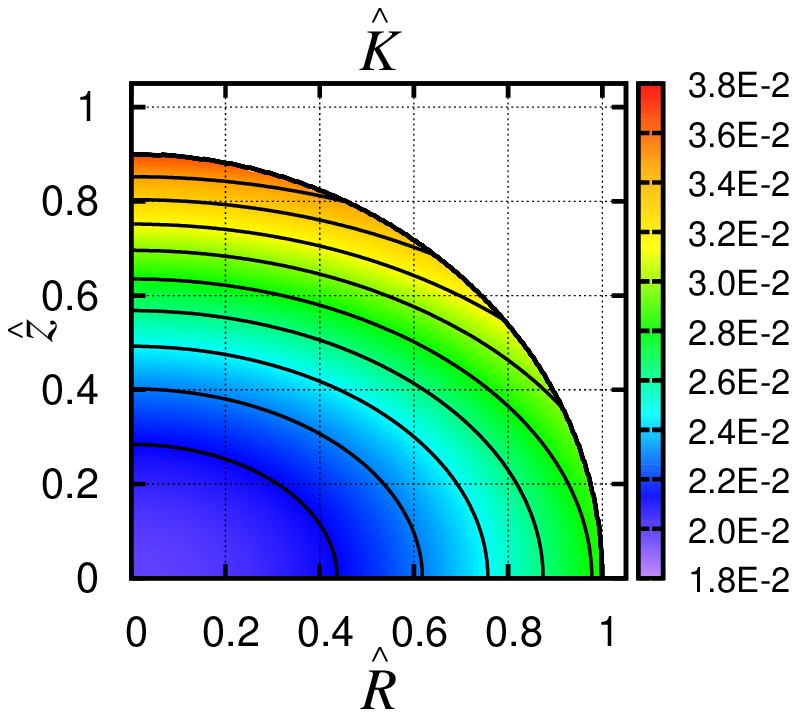}
\includegraphics[width=5.4cm]{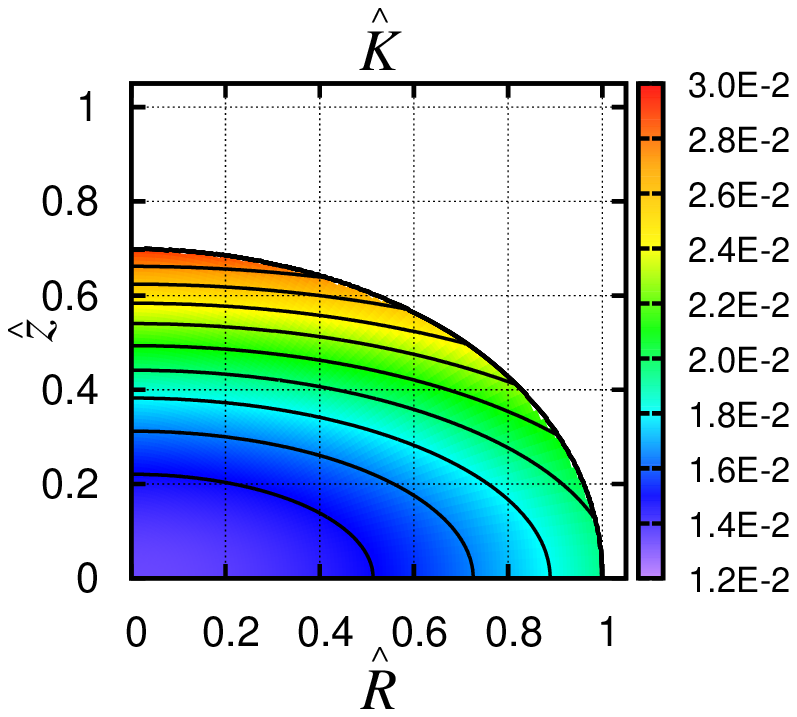}
\includegraphics[width=5.4cm]{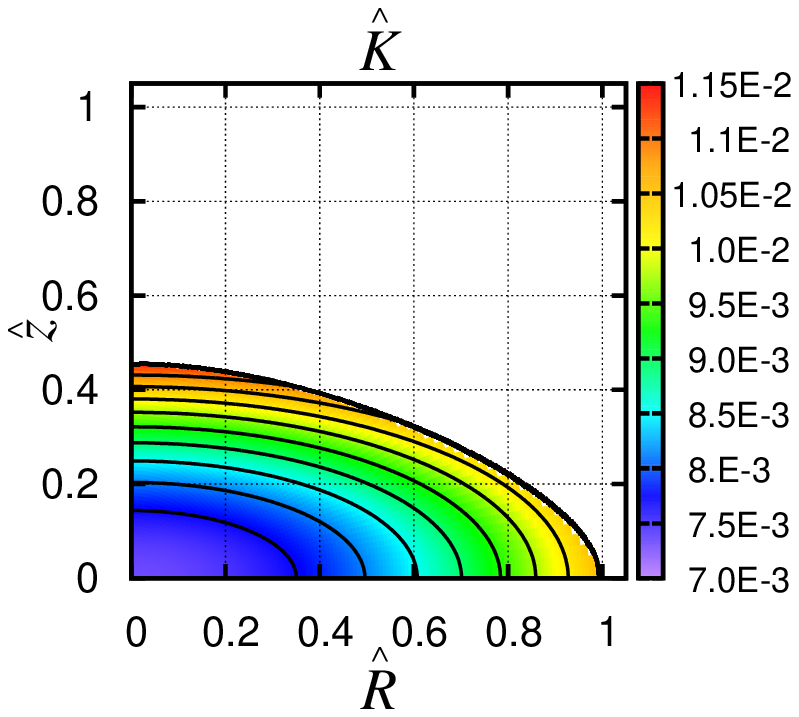}

\includegraphics[width=5.4cm]{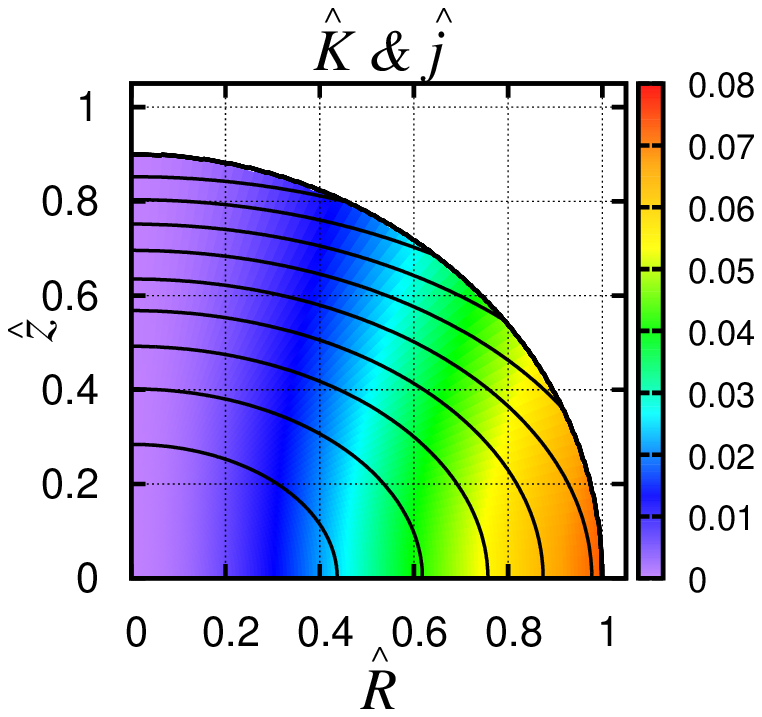}
\includegraphics[width=5.4cm]{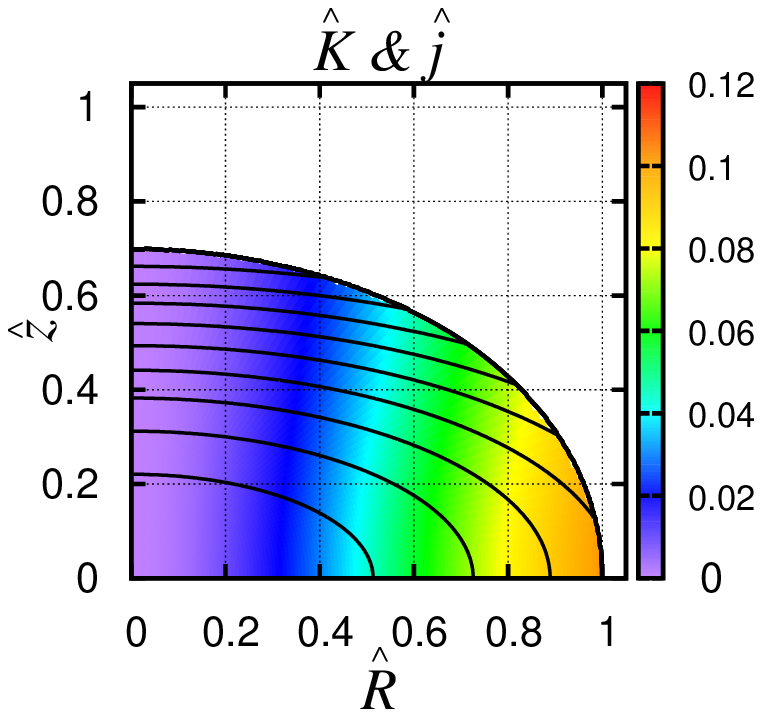}
\includegraphics[width=5.4cm]{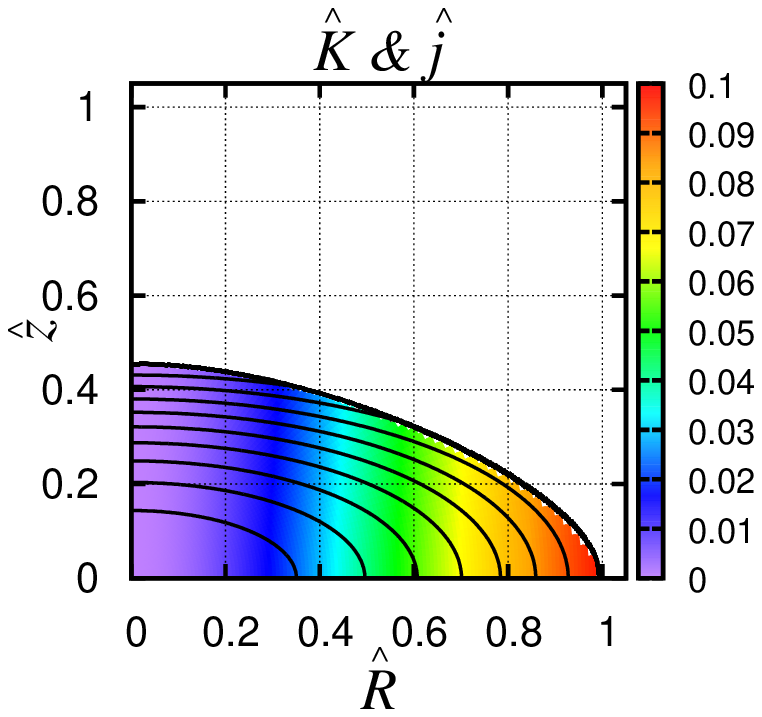}
 
\caption{As per Fig. \ref{Fig:contours1}. 
The three columns of figures display the distributions of solution (o-c) (left-hand panels),
 solution (o-c) and solution of a critical rotation model (o-f), respectively.}
 \label{Fig:contours2}
\end{figure*}


\begin{table*}
 \caption{Physical quantities for the sequence with the oblate isentropic surfaces.
 The solution with '*' denotes the critical rotation model.}
 \begin{tabular}{cccccccccccc}
 \hline
& $q$&  $\hat{\epsilon}$ & $\hat{a}_0$ & $\hat{b}_0$ & $m$ &  $\hat{K}_0$   & $\hat{j}_0^2$ &  $T/|W|$ & $\Pi / |W|$ & VC \\
\hline
(o-a) &  0.900 & 0.45 & 1.0 & 0.65 & 2.0 & 1.965E$-$2 & 2.557E$-$2 & 3.404E$-2$ & 3.106E$-1$ & 8.861E$-6$ \\ 
(o-b) &  0.801 & 0.45 & 1.0 & 0.50 & 2.0 & 1.609E$-$2 & 4.343E$-$2 & 6.365E$-2$ & 2.909E$-1$ & 9.212E$-6$ \\ 
(o-c) &  0.699 & 0.45 & 1.0 & 0.43 & 2.0 & 1.336E$-$2 & 5.253E$-$2 & 8.812E$-2$ & 2.746E$-1$ & 9.698E$-6$ \\ 
(o-d) &  0.599 & 0.45 & 1.0 & 0.41 & 2.0 & 1.118E$-$2 & 5.613E$-$2 & 1.076E$-1$ & 2.616E$-1$ & 1.051E$-6$ \\ 
(o-e) &  0.500 & 0.45 & 1.0 & 0.40 & 2.0 & 8.619E$-$3 & 5.366E$-$2 & 1.202E$-1$ & 2.532E$-1$ & 1.295E$-6$ \\ 
(o-f) &  0.455*& 0.45 & 1.0 & 0.40 & 2.0 & 7.324E$-$3 & 4.860E$-$2 & 1.154E$-2$ & 2.563E$-1$ & 1.573E$-6$ \\ 
\hline
 \end{tabular}
 \label{Tab:tab3}
\end{table*}

Next, a rotating baroclinic  star was calculated with oblate isentropic surfaces.
The parameters of the equation of the state were fixed 
as $\hat{a}_0 > \hat{b}_0$ in order to obtain the oblate 
isentropic surfaces inside the star. A solution sequence 
with $m=2$ was calculated by changing the value of $q$, 
and five solutions (solutions (o-a)--(o-e)) and 
a critical rotation solution (solution (o-f)) were obtained in
the case of the oblate isentropic models.
The numerical results are tabulated in table \ref{Tab:tab3}. 
The structures of the solutions and distributions of 
the physical quantities are displayed in Fig. \ref{Fig:contours2}.
Solution (o-a) (left-hand column), (o-c) (central column) and (o-f) (right-hand column) 
are displayed in Fig. \ref{Fig:contours2}. 
The panels in Fig. \ref{Fig:contours2} are presented similarly to Fig. \ref{Fig:contours1}.
 The arbitrary isopycnic surfaces and isopycnic surfaces of solution (o-a) are also 
 displayed the right-hand panel of \ref{Fig:iso_surfaces} to check the baroclinicity easily. 

 As seen in the isobaric surfaces (solid curve) and the isopycnic surfaces (dashed curves) 
in the right-hand panel of Fig. \ref{Fig:iso_surfaces}
the isopycnic surfaces are always more oblate than the isobaric surfaces.
This tendency is also found in the first row in Fig. \ref{Fig:contours2}
(particularly solution (o-c)).
This is the opposite situation to the spherical isentropic models 
 seen in the previous subsection  because the entropy distributions near 
the stellar surfaces are different from each other.
The third row shows the $\hat{K}$ isosurfaces (contours) and the values of $\hat{K}$ (colour maps). 
The specific entropy in these models also increases outward, 
but the equatorial specific entropy is lower than the polar specific entropy.
These entropy distributions mean that the surface temperature is hotter 
at the poles than at the equator. This distribution of the specific entropy
 is opposite to that of the spherical isentropic model. 
The distribution of the angular velocity is also opposite to that of
the spherical isentropic model.
The second row in Fig. \ref{Fig:contours2} shows the angular velocity distributions. 
The values of $\partial \hat{\Omega} / \partial \hat{z}$ are always negative 
because of the entropy distributions. 
This tendency is also in contrast to the spherical isentropic model. 
According to the Bjerknes--Rosseland rules, 
the value of $\partial \Omega / \partial z$ is always negative if the
isopycnic surfaces are more oblate than the isobaric surfaces 
and the surface temperature is hotter at the poles than at the equator
(\citealt{Tassoul_1978}).
All of these rotating baroclinic stars also satisfy the rules.
The shapes of the angular velocity isosurfaces  are interesting because
they are clearly shellular-type (Fig. \ref{Fig:contours2}). 
In particular, those of solution (o-a) are almost perfectly shellular ones.
Although their shapes are deformed by the rapid rotations,
solutions (o-c) and (o-f) also have shellular-type rotation.
The fourth row in Fig. \ref{Fig:contours2} shows the $\hat{K}$ isosurfaces (contours) and
the values of the specific angular momentum $\hat{j}$ (colour maps). 
The specific angular momentum on each isentropic surface 
also increases as we move from poles to the equator in these models.

Table \ref{Tab:tab3} shows the solution sequence with the oblate isentropic surfaces. 
The critical rotation model (solution (o-f)) was obtained with the oblate isentropic model.
The axis ratio of solution (o-f) is larger than that of the barotropic critical rotation
and the energy ratio $T/|W|$ is smaller than that of the barotropic one. 
The axis ratio is also larger than those of solutions (s2-f) and (s1-f).
The right-hand panels in Fig. \ref{Fig:contours2} show the critical rotation model (solution (o-f)).
The shellular-type rotation is realized even if the stellar rotation is as rapid as critical rotation.

Summarizing the above, these rotating baroclinic stars have oblate isentropic surfaces,
and the isopycnic surfaces are more oblate than their isobaric surfaces 
due to the entropy distributions. Such baroclinicity makes angular velocity similar to that observed 
 for shellular-type rotation, and the value of $\partial \Omega / \partial z$ always
negative inside the star. This condition is also one of 
the Bjerknes--Rosseland rules (see e.g. \citealt{Tassoul_1978}).  
Therefore, all of these rotating baroclinic stars satisfy the Bjerknes--Rosseland rules.

\section{Discussion and summary}

\subsection{Stability of rotating baroclinic stars}

Stability analysis is an important problem
to investigate for  realistic rotating stars. 
Many types of instabilities are known
to occur in rotating baroclinic stars, but 
we restrict ourselves to the consideration of instabilities 
arising from the stellar rotation in this paper. 
In particular, we focus on the dynamical instability due to the rotation 
that occurs because the time-scales of the shear instability arising from 
the differential rotation and the thermal instability (e.g. \citealt{Goldreich_Schubert_1967}) 
are much larger than that of the dynamical instability. 
We consider stability criteria derived from linear perturbation
theory for rotating baroclinic stars.

The stability criteria of rotating stars are characterized by the 
specific angular momentum ($j = \Omega R^2$) distributions.
According to the H{\o}iland criterion, 
a baroclinic star in permanent rotation is dynamically stable with respect to axisymmetric 
 motions if and only if the two following conditions are satisfied: 
(i) the specific entropy $s$ never decreases outward, and 
(ii) on each surface $s =$ constant (specific entropy constant), 
the angular momentum per unit mass (specific angular momentum $j$) increases 
from the poles to the equator (\citealt{Tassoul_1978}).
 
The thermal structures are totally independently and freely introduced 
to the mechanical structures in Newtonian gravity, 
because the thermal structures are obtained by assuming and adding 
thermal balance equations. If the ideal gas is adopted 
as one of the simplest equation of state, 
the specific entropy is determined and becomes proportional to $\hat{K}$. 
As seen earlier in this paper,  the specific entropy never increases 
outward in all of the rotating baroclinic solutions in this paper. 
Therefore, all of 
the rotating baroclinic stars in this paper satisfy condition (i).
The profiles of $j$ on $s = $ constant surfaces ($\hat{K}$ constant surfaces) 
are displayed in bottom panels in Figs. \ref{Fig:contours1} and \ref{Fig:contours2}.
As shown in these panels, the specific angular momentum on isentropic surface 
never decreases in both barotropic models. All of the rotating baroclinic stars in this paper also 
satisfy condition (ii). Therefore all of the rotating baroclinic stars satisfy the H{\o}iland criterion.
They are dynamically stable configurations against 
dynamical instability due to rotation. Further analysis of baroclinic rotating
stars with realistic equations of state would be interest, 
but thermal balance (energy equation) must be solved 
to obtain the baroclinic stars with realistic equation of state.
It is beyond the scope of the present paper. 
We will construct realistic rotating baroclinic 
models using this new numerical method and systematically analyse their stability in future works.
 

\subsection{Shellular rotation}

Many stellar evolution codes for rotating stars have adopted 
shellular rotation as one of their rotation models 
(e.g. \citealt{Meynet_Maeder_1997}; \citealt{Talon_et_al_1997}; 
\citealt{Heger_Langer_Woosley_2000}; \citealt{Mathis_Palacios_Zahn_2004}; \citealt{Maeder_Meynet_2005}; 
\citealt{Potter_Tout_Brott_2012,Potter_Tout_Eldridge_2012}; \citealt{Takahashi_Umeda_Yoshida_2014}). 
The first shellular rotation model was formulated and derived by \cite{Zahn_1992} using a 
perturbative approach.  When the turbulence 
is anisotropic with stronger matter transport in the horizontal directions than in the vertical,
shellular rotation is realized (\citealt{Zahn_1992}).
As a result, the horizontal-dependence of the angular velocity disappears and 
the angular velocity depends on only the distance from the centre of the star.
This kind of shellular rotation was adopted as an initial rotation profile 
of core-collapse supernova simulations (e.g. \citealt{Yamada_Sato_1994}).  
However, nobody had successfully  investigated self-consistent configurations of 
rapidly rotating stars with shellular rotation systematically because of 
the difficulty of obtaining rotating baroclinic stars. 

Some previous works showed self-consistent rotating baroclinic stars with
non-cylindrical rotations (\citealt{Uryu_Eriguchi_1994}; \citealt{Uryu_Eriguchi_1995}; 
\citealt{Espinosa_Rieutord_2007}; \citealt{Espinosa_Rieutord_2013}),
but their rotation distributions were far from shellular.
On the other hand, \cite{Roxburgh_2006} calculated rotating baroclinic stars by fixing
the distribution of the shellular rotation,
but neither calculated solutions systematically
nor obtained critical rotation models. \cite{Kiuchi_Nagakura_Yamada_2010} 
investigated multi-layered configurations
in differentially rotational equilibrium. While they did obtain
rapidly rotating solutions systematically, they calculated rotating barotropic stars only.

In contrast, this paper has investigated rapidly rotating baroclinic stars systematically 
as in Section 3.3.  The oblate isentropic 
surfaces models were found to have shellular-type angular velocity distributions. 
The equatorial surface temperature of the rotating baroclinic star 
is hotter than the polar temperature. These kinds of entropy and temperature 
distributions generate shellular-type rotation. 
Shellular-type rotation is realized even if the star has very rapid rotation. 
Solution (o-f) is a critical rotation model and its distribution of the 
angular velocity is shellular-type (Fig.\ref {Fig:contours2}). 
These are the first self-consistent and systematic solutions of the 
rapidly baroclinic stars with shellular-type rotations.
These shellular-type rotating stars are dynamically stable 
as they all  satisfy the H{\o}iland criterion as seen in Section 4.1.
This kind of shellular-type rotation would be realized in many 
rapidly rotating   stars.

\subsection{Summary}

A versatile numerical method for 
obtaining structures of rapidly rotating baroclinic stars was investigated.
The novel numerical method was based on the self-consistent field scheme and 
the solution was obtained iteratively.
Both rotating  barotropic and baroclinic stars 
were calculated systematically by using the new method. 
The accuracy of the new method was verified by checking the convergence
of numerical results. The relative values of the virial relation were small and converged well.
Further, the solutions from the new method and those obtained by the
HSCF scheme, a well-established numerical scheme, were compared, 
and the solutions were very similar.

Two types of rotating baroclinic stars
--- spherical isentropic surfaces models and 
oblate isentropic surfaces models ---
were investigated by fixing functional forms of a simple baroclinic equation of 
state. Solution sequences of both models were calculated systematically
and the critical rotation models were also obtained.
All of the rotating baroclinic stars satisfy the Bjerknes--Rosseland condition.
 In the spherical isentropic surfaces model, 
the isobaric surfaces are more oblate than the isopycnic surfaces and the
values of $\partial \Omega / \partial z$ are always positive throughout the star.
In the oblate isentropic surfaces model, on the other hand,
the isopycnic surfaces are more oblate than the isobaric surfaces and the 
values of $\partial \Omega / \partial z$ are always negative throughout the star.
The baroclinic star with shellular-type rotation was found to be realized when 
isentropic surfaces are oblate and 
the equatorial surface temperature of the rotating baroclinic star 
is hotter than the polar temperature if it is assumed that the star is the ideal gas star.
The shellular-type rotation is realized even if the rotation is as rapid as possible. 
These are the first self-consistent and systematic solutions of the
rapidly rotating baroclinic stars with shellular-type rotations.

The H{\o}iland criterion of the rotating baroclinic stars was checked, with
all  solutions in this paper satisfying the criterion. 
Therefore, these rotating baroclinic star are stable against the dynamical 
instability due to the rotation.  Realistic rotating baroclinic stars with a
realistic equation of state are needed to investigate the evolution of the
rapidly realistic rotating stars. Further works may  extend the numerical method and calculate 
realistic rotating baroclinic stars.

\section*{Acknowledgments}

The author would like to thank N. Yasutake, S. Yamada, T. Urushibata and Y. Eriguchi 
for valuable discussions and useful comments.
The author would also like to thank the anonymous reviewer for useful suggestions.
The author is supported by Grant-in-Aid for Scientific Research on Innovative Areas, 
No.24103006.

\bibliographystyle{mn}


\appendix

\section{Details of numerical scheme}

 The numerical scheme of the new method is described in detail.
The numerical scheme is based on a self-consistent field iteration scheme 
(\citealt{Ostriker_Mark_1968}; \citealt{Hachisu_1986a,Hachisu_1986b}).
The gravitational field and all physical quantities are determined iteratively. 
 A numerical domain of the computation is defined as $0 \leq \hat{r} \leq 2 $ in the
 radial direction and $0 \leq \theta \leq \pi/2$ in the angular direction. 
 The numerical domains are discretized into mesh points with 
 equal intervals $\Delta\hat{r}$ and $\Delta \theta$, respectively as
 \begin{eqnarray}
  \hat{r}_i = i \times \Delta \hat{r}, \hspace{10pt} \Delta \hat{r} = \frac{1}{N_r}, \hspace{10pt} i = 0, 1, 2, 3, \cdots,
   \label{Eq:app1}
 \end{eqnarray}
 \begin{eqnarray}
  \theta_j = (j-1) \times \Delta \theta, \hspace{10pt} \Delta \theta = \frac{\pi/2}{N_\theta - 1}, \hspace{10pt}
 j = 1, 2, 3, \cdots.
   \label{Eq:app2}
 \end{eqnarray}
Physical variables are defined at the cross point of two mesh lines, i.e.,
\begin{eqnarray}
 \hat{\rho}_{i,j} = \hat{\rho}(\hat{r}_i, \theta_j).
\end{eqnarray}

Before the first iteration cycle, 
a one-dimensional spherical non-rotating barotropic star is used 
as the initial guess for two-dimensional rotating star. 
The value of $\hat{K}_0$ is determined to 
make the radius of the 1D spherical star unity.
The values of $\hat{K}_0$ and the 1D star are used for the iteration.

We start the iteration cycle after the initial guesses are obtained. 
First, we calculate the gravitational potential $\hat{\phi}$ from the
initial guess of $\hat{\rho}$ as
\begin{eqnarray}
 &&\hat{\phi} = - \sum_{\ell=0}^{\ell_{\max}} P_{2\ell} (\cos \theta)
\int_0^{\infty} \hat{r}'^2 f_{2\ell}(\hat{r}, \hat{r}') d\hat{r}' 
\nonumber \\
&& \times \int_0^{\pi/2} P_{2\ell}(\cos \theta') \sin \theta' d\theta' 
\hat{\rho}(\hat{r}', \theta'),
\end{eqnarray}
where $P_{2\ell}$ is a Lgendre polynomial and $f_{2\ell}(\hat{r},\hat{r}')$ is a function defined by
\begin{eqnarray}
 f_{2\ell}(\hat{r}, \hat{r}') = \left\{
	      \begin{array}{cc}
	       \hat{r}'^{2\ell} / \hat{r}^{2\ell + 1} & (\hat{r}  > \hat{r}') \\
	       \hat{r}^{2\ell}  / \hat{r}'^{2\ell + 1} & (\hat{r} < \hat{r}') .
	      \end{array}
\right.
\end{eqnarray}
Simpson's scheme is used to calculate the integrals numerically.
$\ell_{\max}$ is fixed as $2 \ell_{\max} = 40$ in all numerical calculations in this paper.

Next, we calculate the values of $\hat{j}_0$ and $\hat{K}_0$.
These values are determined in order to make the 
equatorial radius unity and the axis ratio $q$.
First, we calculate the value of $\hat{K}_0$ by
solving the dimensionless form of the $r$ component 
of the Euler equation (equation \ref{Eq:eular_r}) with the equation of state (equation \ref{Eq:p_k_rho2}) 
on the rotational axis. The equations on the rotational axis are respectively discretized as follow: 
\begin{eqnarray}
 \frac{\hat{p}_{i, 1} - \hat{p}_{i-1,1}}{\Delta \hat{r}} = \frac{(\hat{\rho}_{i,1} + \hat{\rho}_{i-1,1})}{2} 
\frac{(\hat{\phi}_{i,1} - \hat{\phi}_{i-1,1})}{\Delta \hat{r}},
\label{Eq:dis_eular_r_p}
\end{eqnarray}
\begin{eqnarray}
\hat{p}_{i,1} = \hat{K}_0  \left\{ 1 + \epsilon \left( \frac{1}{\hat{b}_0^2} 
\right) \hat{r}_{i,1}^m \right\} \left( \hat{\rho}_{i,1} \right)^{1 + \frac{1}{N}}.
\label{Eq:dis_EOS1_p}
\end{eqnarray}
A central difference was adopted in this numerical scheme.
We start the integration at the centre of the star ($\hat{r} = 0$). Since we have defined 
the dimensionless form of the central density in equations (\ref{Eq:rho_c}) and (\ref{Eq:p_c}),
the values of $\hat{\rho}_{1,1}$ and $\hat{p}_{1,1}$ are obtained by 
solving the discretized equations using  the values of $\hat{\rho}_c$ and $\hat{p}_c$. 
Then, the values of $\hat{\rho}_{2,1}$ and  $\hat{p}_{2,1}$ are determined using the values of
$\hat{\rho}_{1,1}$ and $\hat{p}_{1,1}$. The integration is continued until the density $\hat{\rho}_{i,1}$ becomes 
zero ($\hat{\rho}_{i,1} = 0)$,  and we obtain the value 
of the polar radius $\hat{r}_{\rm pol.}$ after the integration.
If the value of $\hat{r}_{\rm pol.}$ is different from the value of $q$, 
we change the value of $\hat{K}_0$ and restart the integration 
using a shooting-type method (\citealt{Numerical_Recipes})
until we obtain the value of $\hat{K}_0$ that makes the polar radius ($\hat{r}_{\rm pol.}$) $q$.
Next, we calculate the value of $\hat{j}_0$ by solving the dimensionless form of the $r$ component 
of the Euler equation (equation \ref{Eq:eular_r}) and the equation of state (equation \ref{Eq:p_k_rho2})
on the equatorial axis. The equations on the equatorial plane are discretized as follows:
\begin{eqnarray}
&& \frac{\hat{p}_{i, N_\theta} - \hat{p}_{i-1,N_\theta}}{\Delta \hat{r}} 
= \frac{(\hat{\rho}_{i,N_\theta} + \hat{\rho}_{i-1,N_\theta})}{2} 
\frac{(\hat{\phi}_{i,N_\theta} - \hat{\phi}_{i-1,N_\theta})}{\Delta \hat{r}} \nonumber \\
&&+ \frac{(\hat{r}_{i} + \hat{r}_{i-1} )}{2} \frac{(\hat{\rho}_{i,N_\theta} + \hat{\rho}_{i-1,N_\theta} )}{2}
\frac{(\hat{\Omega}^2_{i,N_\theta} + \hat{\Omega}^2_{i-1,N_\theta} )}{2},
\label{Eq:dis_eular_r_e}
\end{eqnarray}
\begin{eqnarray}
 \hat{p}_{i,N_\theta} = \hat{K}_0  \left\{ 1 + \epsilon \left( \frac{1}{\hat{a}_0^2}  
\right) \hat{r}_{i,N_\theta}^m \right\} \left( \hat{\rho}_{i,N_\theta} \right)^{1 + \frac{1}{N}}.
\label{Eq:dis_EOS1_e}
\end{eqnarray}
The value of $\hat{j}_0$ is determined in the same manner to make the equatorial radius unity.

\begin{figure*}
\includegraphics[width=8cm]{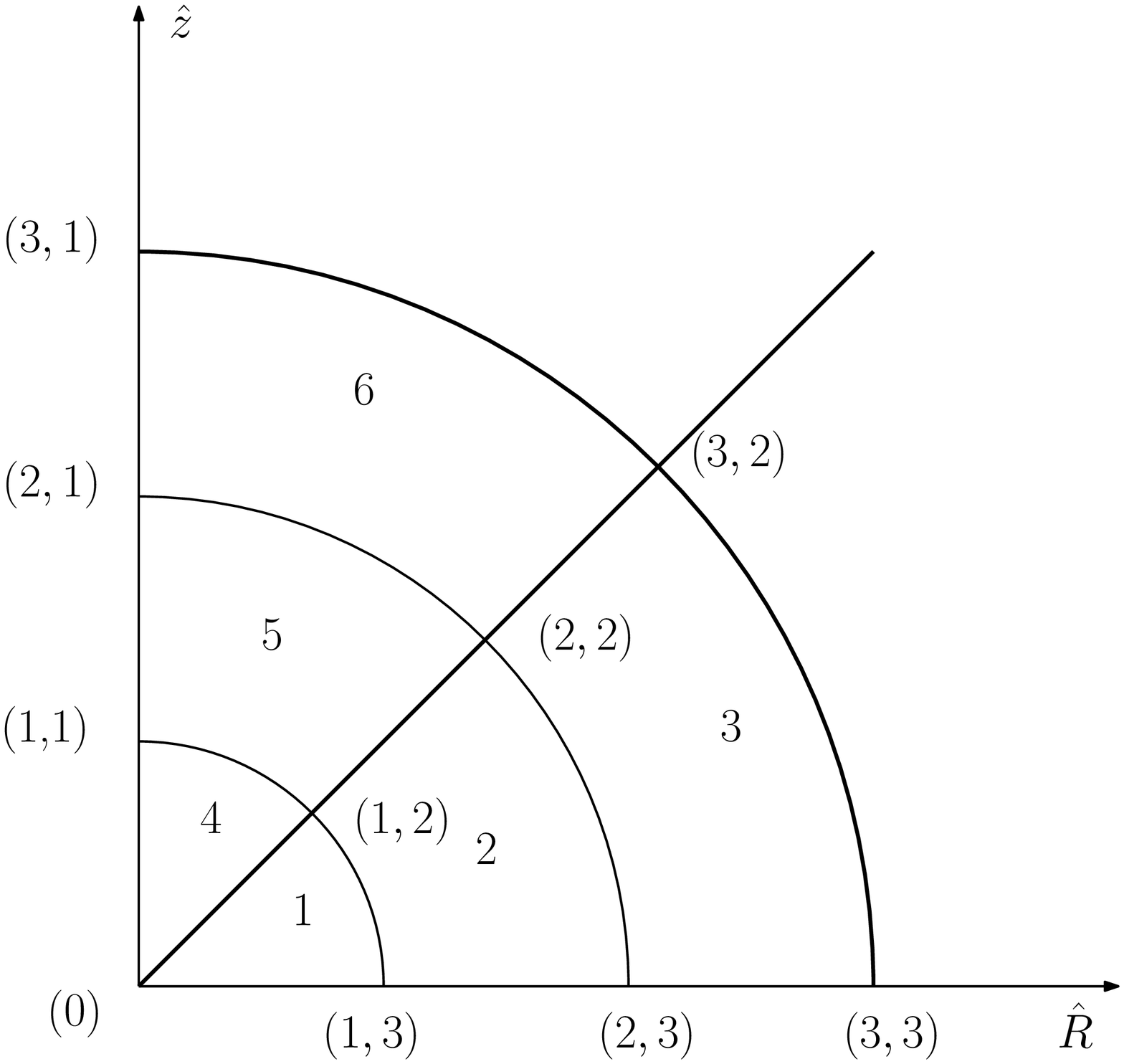}
\includegraphics[width=8cm]{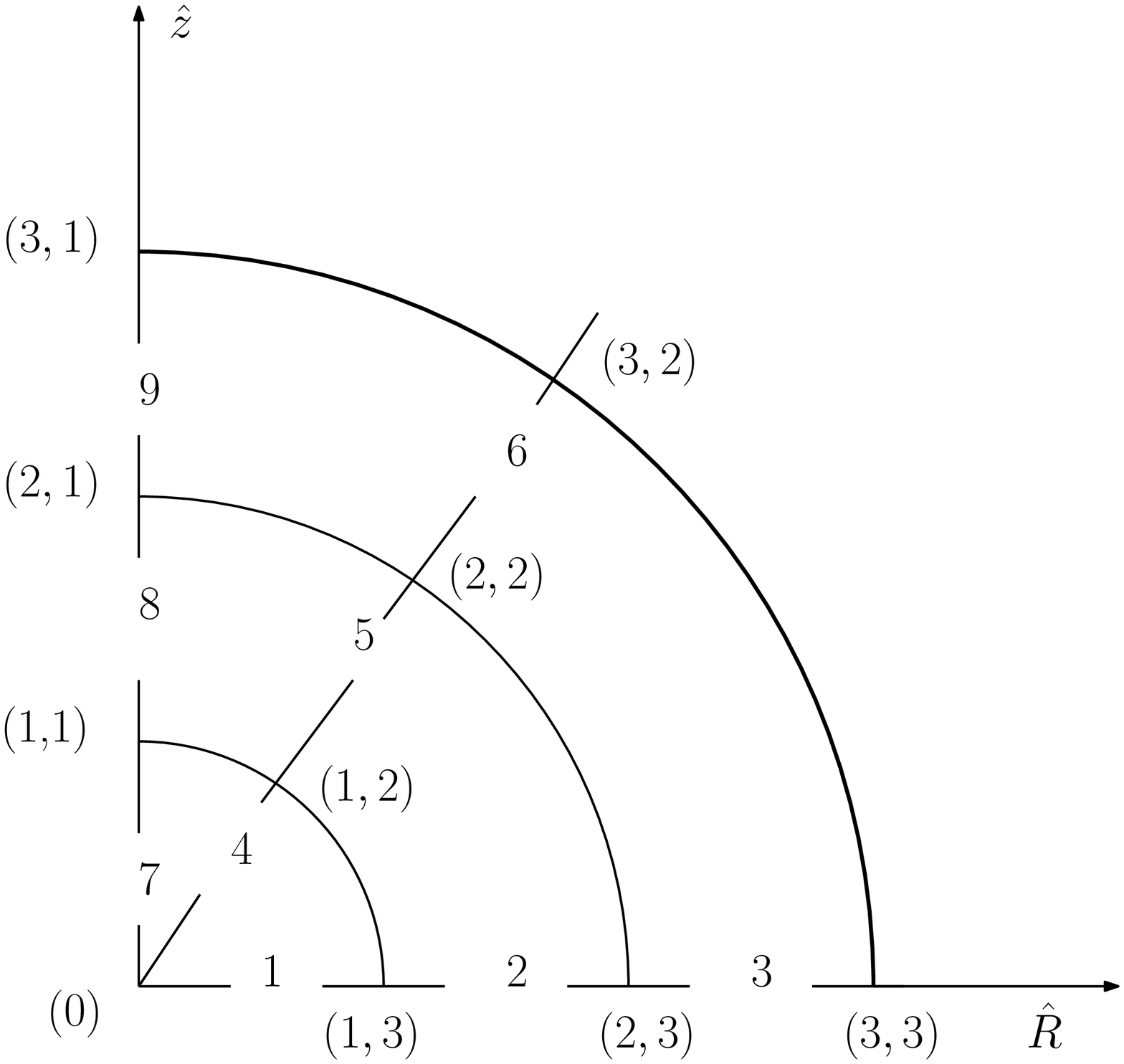}
\caption{Schematic images of the numerical scheme for $\hat{\Omega}$ (left-hand panel)
and $\hat{\rho}$, $\hat{p}$ (right-hand panel).
The thick curve denotes the stellar surface. 
$(i,j)$ denotes the mesh point and each number refers to the sequence of the integration. 
We start the integration of equations at point 1.}
\label{Fig:fig}
\end{figure*}

After the new values of $\hat{K}_0$ and  $\hat{j}_0$ have been obtained, 
the distribution of the angular velocity can be calculated 
using the dimensionless forms of equations (\ref{Eq:baroclinicity}) and (\ref{Eq:boundary2}). 
Four neighbouring mesh points
[$(i-1,j)$, $(i,j)$, $(i-1,j+1)$ and $(i, j+1)$] are used to solve equation (\ref{Eq:baroclinicity})
in this numerical scheme.  The equation is  
discretized at the centre of the four mesh points (left-hand panel of Fig. \ref{Fig:fig}). 
The explicit form of the discretized equation is
\begin{eqnarray}
 && \hat{r}_m^2 \sin \theta_m \cos \theta_m 
 \frac{(\hat{\Omega}^2_{i,j} + \hat{\Omega}^2_{i,j+1}) - (\hat{\Omega}^2_{i-1,j} + \hat{\Omega}^2_{i-1,j+1})}{2\Delta \hat{r}}
\nonumber \\
&& - \hat{r}_m \sin^2 \theta_m \frac{(\hat{\Omega}^2_{i,j+1} + \hat{\Omega}^2_{i-1,j+1}) - (\hat{\Omega}^2_{i,j}+\hat{\Omega}^2_{i-1,j}) }{2\Delta \theta}
\nonumber \\
&=& \frac{1}{\hat{\rho}_m^2} 
\Bigg[
\left\{
\frac{(\hat{\rho}_{i,j+1} + \hat{\rho}_{i-1, j+1}) - (\hat{\rho}_{i,j} + \hat{\rho}_{i-1,j} )}{2\Delta \theta} 
\right\}
\nonumber \\
&&\times
\left\{
\frac{(\hat{p}_{i,j+1} + \hat{p}_{i, j}) - (\hat{p}_{i-1,j+1} + \hat{p}_{i-1,j} )}{2\Delta \hat{r}}
\right\} 
\nonumber \\
&-&
\left\{
\frac{(\hat{p}_{i,j+1} + \hat{p}_{i-1, j+1}) - (\hat{p}_{i,j} + \hat{p}_{i-1,j} )}{2\Delta \theta} 
\right\}
\nonumber \\
&& \times
\left\{
\frac{(\hat{\rho}_{i,j+1} + \hat{\rho}_{i, j}) - (\hat{\rho}_{i-1,j+1} + \hat{\rho}_{i-1,j} )}{2\Delta \hat{r}}
\right\}
\Bigg],
\label{Eq:Omega2}
\end{eqnarray}
where $\hat{r}_m$, $\theta_m$ and $\hat{\rho}_m$ are
\begin{eqnarray}
 \hat{r}_m = \frac{1}{2} \left(\hat{r}_{i} + \hat{r}_{i-1} \right),
\end{eqnarray}
\begin{eqnarray}
 \theta_m = \frac{1}{2} \left(\theta_{j+1} + \theta_{j}  \right),
\end{eqnarray}
\begin{eqnarray}
 \hat{\rho}_m = \frac{1}{4} \left( \hat{\rho}_{i-1,j} + \hat{\rho}_{i,j} + \hat{\rho}_{i-1,j+1} + \hat{\rho}_{i,j+1}  \right).
\end{eqnarray}
In this scheme, the value of $\hat{\Omega}_{i,j}^2$ is calculated by using the values of 
$\hat{\Omega}_{i-1,j}^2$, $\hat{\Omega}^2_{i-1,j+1}$ and $\hat{\Omega}^2_{i,j+1}$.
Fig. \ref{Fig:fig} displays schematic images of this scheme.
For simplicity, Fig. \ref{Fig:fig} has assumed mesh numbers $N_r = 3$ and $N_\theta = 3$. 
We start the integration at point 1 in the left-hand panel of Fig. \ref{Fig:fig}.
First, the boundary condition on the equatorial plane is set.
The values of $\hat{\Omega}^2_{0}$, $\hat{\Omega}^2_{1,3}$, 
$\hat{\Omega}^2_{2,3}$ and $\hat{\Omega}^2_{3,3}$ are fixed during the iterations.
At point 1 in the left-hand panel of Fig. \ref{Fig:fig}, 
we obtain the value of $\Omega^2_{1,2}$ from equation (\ref{Eq:Omega2}) by using the 
values of $\hat{\Omega}^2_{0}$ and $\hat{\Omega}^2_{1,3}$ because 
$\hat{\Omega}^2_{0,3} = \hat{\Omega}^2_{0,2} = \hat{\Omega}^2_{0}$ at point 1.
Then, we move to point 2 and calculate $\hat{\Omega}^2_{2,2}$ by 
using the values of  $\hat{\Omega}^2_{1,3}$, $\hat{\Omega}^2_{2,3}$ and $\hat{\Omega}^2_{1,2}$,
and solving the equation. 

Finally, we calculate the distributions of $\hat{\rho}$ and $\hat{p}$ from
equations (\ref{Eq:eular_r}) and (\ref{Eq:p_k_rho2}) by using the distributions 
of $\hat{\phi}$ and  $\hat{\Omega}^2$. 
The two mesh points ($(i,j)$ and $(i-1,j)$) are used in this calculation. 
The equation is  discretized  as
\begin{eqnarray}
&& \frac{\hat{p}_{i, j} - \hat{p}_{i-1,j}}{\Delta \hat{r}} = \frac{(\hat{\rho}_{i,j} + \hat{\rho}_{i-1,j})}{2} 
\frac{(\hat{\phi}_{i,j} - \hat{\phi}_{i-1,j})}{\Delta \hat{r}} 
\nonumber \\ 
&&
+ \frac{(\hat{r}_{i} + \hat{r}_{i-1})}{2} \sin \theta^2_{j}
\frac{(\hat{\rho}_{i,j} + \hat{\rho}_{i-1,j})}{2} \frac{(\hat{\Omega}_{i,j}^2 + \hat{\Omega}^2_{i-1,j})}{2}
\label{Eq:dis_eular_r}
\end{eqnarray}
and the equation of state is given as 
\begin{eqnarray}
 \hat{p}_{i,j} =  \hat{K}_0  \left\{ 1 + \epsilon \left( \frac{\sin \theta_j}{\hat{a}_0^2} 
+ \frac{\cos \theta_j}{\hat{b}_0^2} \right) 
\hat{r}_{i}^m \right\} \left( \hat{\rho}_{i,j} \right)^{1 + \frac{1}{N}}.
\label{Eq:dis_EOS1}
\end{eqnarray}
We start the integration at point 1 in the right-hand panel of Fig. \ref{Fig:fig}.
Since the central density and the central pressure have been fixed as $\hat{\rho}_0 = \hat{\rho}_c = 1$ and 
$\hat{p}_{0} = \hat{p}_c$ by equations (\ref{Eq:p_c}) and (\ref{Eq:dis_EOS1}),
we  calculate $\hat{\rho}_{1,3}$ and $\hat{p}_{1,3}$ by equation (\ref{Eq:dis_eular_r}).
Next, we move to point 2 and obtain $\hat{\rho}_{2,3}$ and $\hat{p}_{2,3}$ 
by using the values of $\hat{\rho}_{1,3}$ and $\hat{p}_{1,3}$ and equations 
(\ref{Eq:dis_eular_r}) and (\ref{Eq:dis_EOS1}).
We can obtain the values of $\hat{\rho}_{i,j}$ and $\hat{p}_{i,j}$ 
by using the values of $\hat{\rho}_{i-1,j}$ and $\hat{p}_{i-1,j}$ .
 If the value of $\hat{\rho}_{i,j}$ or $\hat{p}_{i,j}$ is zero, 
we search the location of the stellar surface $\hat{r}_s (\theta)$,
where $\hat{r}_s(\theta)$ denotes the dimensionless radius in each $\theta$ direction.
After calculating the location of the surface, we move to point 4 and 
continue the integration. After the integration on the rotational axis (point 6 in the panel), 
we have obtained new distributions of $\hat{\rho}$ and $\hat{p}$ and
one iteration cycle is finished.

In summary, the numerical iteration cycle of this method is as follows.
\begin{enumerate}
 \item Set the parameters ($N$, $q$) and fix the 
       functional forms of $\hat{\Omega}^2$ on the equatorial plane and the equation of state.
 \item Make an initial guess for the two-dimensional rotating baroclinic star.
       Calculate a one-dimensional spherical barotropic star for initial guesses of the iteration. 
 \item Calculate the gravitational potential $\hat{\phi}$ by 
       using the distribution of $\hat{\rho}$.
 \item Calculate the values of $\hat{K}_0$ and $\hat{j}_0$ via a shooting-type method 
       in order to make the equatorial radius unity and the axis ratio $q$.
 \item Calculate the angular velocity $\hat{\Omega}$ by using the distribution 
       of $\hat{\phi}$ obtained in (iii).
 \item Solve the $r$-component of the Euler equation with the equation of state
       and obtain the new distributions of density $\hat{\rho}$ and pressure $\hat{p}$ 
       by using the distributions of $\hat{\phi}$ obtained in (iii)  
       and $\hat{\Omega}$ obtained in (v).
 \item Check the convergence of the system. 
       Compare the new and old density distributions.
       If the relative differences between the new distributions and the old distributions are
       sufficiently small (e.g. smaller than $\sim 10^{-4}$), the system is considered to converge well
       and the iteration cycle finishes as the rotating baroclinic star has been obtained in equilibrium.
       If the system does not converge, return to (iii) and start a new iteration cycle until the 
       system converge.
\end{enumerate}

\end{document}